\documentclass[twocolumn,superscriptaddress,aps,prx,longbibliography]{revtex4-1}
\usepackage{amsmath,amssymb}
\usepackage{graphicx}
\usepackage{epsfig}
\usepackage{hyperref}
\usepackage{xcolor,colordvi}
\usepackage{float}

\usepackage{tikz}
\usepackage{xcolor}
\usepackage{subcaption}
\usepackage{ragged2e}

\usepackage{lmodern}
\usetikzlibrary{decorations.pathmorphing}

\tikzset{snake it/.style={decorate, decoration=snake}}

\usetikzlibrary{arrows}

\usetikzlibrary{decorations.markings}

\newcommand{\be}{\begin{equation}}
\newcommand{\ee}{\end{equation}}
\newcommand{\bea}{\begin{eqnarray}}
\newcommand{\eea}{\end{eqnarray}}
\newcommand{\ba}{\begin{eqnarray*}}
\newcommand{\ea}{\end{eqnarray*}}

\usepackage{bbold}

\begin{document}

\title{Fast Scrambling at the Boundary}
\author{Ancel Larzul}
\affiliation{JEIP, UAR 3573 CNRS, Coll\`{e}ge de France, PSL Research University, 11 Place Marcelin Berthelot, 75321 Paris Cedex 05, France}
\author{Anirvan M. Sengupta}
\affiliation{Department of Physics and Astronomy, Rutgers University, Piscataway, New Jersey 08854, USA}
\affiliation{Center for Computational Quantum Physics, Flatiron Institute, New York, New York 10010, USA}
\affiliation{Center for Computational Mathematics, Flatiron Institute, New York, New York 10010, USA}
\author{Antoine Georges}
	\affiliation{Center for Computational Quantum Physics, Flatiron Institute, New York, New York 10010, USA}
    \affiliation{Coll\`ege de France, PSL University, 11 place Marcelin Berthelot, 75005 Paris, France}	
	\affiliation{Department of Quantum Matter Physics, University of Geneva, 24 quai Ernest-Ansermet, 1211 Geneva, Switzerland}
	\affiliation{CPHT, CNRS, Ecole Polytechnique, IP Paris, F-91128 Palaiseau, France}
\author{Marco Schir\`o}
\affiliation{JEIP, UAR 3573 CNRS, Coll\`{e}ge de France, PSL Research University, 11 Place Marcelin Berthelot, 75321 Paris Cedex 05, France}

\begin{abstract}
Many-body systems which saturate the quantum bound on chaos are attracting interest across a wide range of fields. Notable examples include the Sachdev-Ye-Kitaev model and its variations, all characterised by some form or randomness and all to all couplings. Here we study many-body quantum chaos in a quantum impurity model showing Non-Fermi-Liquid physics, the overscreened multichannel $SU(N)$ Kondo model. We compute exactly the low-temperature behavior of the out-of time order correlator in the limit of large $N$ and large number of channels $K$, at fixed ratio $\gamma=K/N$. Due to strong correlations at the impurity site the spin fractionalizes in auxiliary fermions and bosons. We show that all the degrees of freedom of our theory acquire a Lyapunov exponent which is linear in temperature as $T\rightarrow 0$, with a prefactor that depends on $\gamma$. Remarkably, for $N=K$ the impurity spin displays maximal chaos, while bosons and fermions only get up to half of the maximal Lyapunov exponent. Our results highlights two new features: a non-disordered model which is maximally chaotic due to strong correlations at its boundary and a fractionalization of quantum chaos.
\end{abstract}

\maketitle

\section{Introduction}

Unitarity of quantum mechanics implies that information initially encoded in local degrees of freedom cannot be lost under time evolution but only scrambled into non-local ones, inaccessible to external probes~\cite{PatrickHayden_2007,sekino2008fast,shenker2014blackholes,hosur2016chaos}. The ability of quantum many-body systems to act as scramblers has emerged in recent years as a new lens to classify phases of matter. A powerful proxy to quantify quantum scrambling is provided by out-of-time ordered correlators (OTOC)~\cite{larkin1969semiclassical,Aleiner2016microscopic,swingle2018unscrambling,xu2024scrambling} originally introduced in the quasi-classical theory of superconductivity and more recently extended to a variety of different settings, from field theories~\cite{stanford2016manybody,chowdhury2017onset,patel2017quantumbutterfly} to random circuits~\cite{nahum2018operator,khemani2018operator,chan2018solution,kos2018manybody,vonkeyserlingk2018operator} to models with local bounded Hilbert space~\cite{kukuljan2017weak}. Whenever the early-time growth of the OTOC is exponential in time one can define a quantum Lyapunov exponent $\lambda_L$, which generalizes the butterfly effect to the quantum domain and it is often taken as a measure of many-body quantum chaos. Quantum mechanics fundamentally influences this exponent by setting an upper bound on chaos~\cite{maldacena2016abound,tsujii2018bound,pappalardi2022quantum}, set essentially by the temperature of the system, $\lambda_L\le 2\pi k_BT/\hbar$. Systems which saturate this bound, so called fast scramblers, such as the celebrated Sachdev-Ye-Kitaev (SYK) model~\cite{sachdev93gapless,parcollet99nonfermi,georges2001quantum,kitaevtalk,kitaev2018softmode}, are therefore particularly interesting for a broad range of communities~\cite{Maldacena_2016,polchinski2016spectrum,jensen2016chaos,facoetti2019classical,bera2022quantum}.

A natural place to look for fast scramblers and maximal chaos are quantum phases of matter without well-defined quasiparticles. These violate Landau's Fermi Liquid paradigm and display unusual transport properties, which have attracted continuous interest over the past decades~\cite{chowdhury2022sachdev}. Recently a fresh new view on these Non-Fermi-Liquids has emerged which is rooted in their ability to thermalize in the fastest possible way~\cite{sachdev2015bekenstein,hartnoll2018holographic,hartnoll2022colloquium} and to efficiently (maximally) scramble quantum information~\cite{sekino2008fast,shenker2014blackholes}. 
Understanding the chaotic properties of Non-Fermi Liquids has become therefore an urgent problem and recent literature has been devoted to the study of OTOC for different types of models, including variations of the SYK model~\cite{gu2017local,banerjee2017solvable,guo2019transport,davis2023quantum} or interacting fermions coupled to critical bosons~\cite{patel2017quantum,kim2021dirac,tikhanovskaya2022maximal}. 
A common thread in the search for Non-Fermi Liquids and maximally chaotic quantum matter is the presence of quenched randomness in the form of all to all couplings, which are solvable~\cite{patel2023universal}. Whether randomness is a fundamental requirement is however not understood. This raises the question of what is the simplest model displaying maximal chaos. In this work we show that, quite surprisingly,
randomness is not needed for maximal chaos.
Indeed we will be able to display a model for maximal chaos which only describes a quantum spin strongly coupled to bath, i.e. a quantum impurity.

Quantum impurity models have been long known to display exotic Non-Fermi Liquid behavior~\cite{nozieres1980kondo,cox1998exotic}, whenever for example a single magnetic impurity possibly with non-trivial orbital character interacts with multiple channels of conduction electrons~\cite{andrei1984solution,affleck1991universal} or in presence of magnetic interactions between impurities~\cite{jayaprakash1981twoimpurity,jones1988lowtemperature,georges1995solution,georges1997kondo}. The two-channels Kondo model in this respect is among the most studied example of NFL behavior, also experimentally relevant for mesoscopic physics~\cite{matveev1995coulomb,potok2007observation,mebrahtu2012quantum,iftikhar2015two,iftikhar2018tunable}. 
Here the break-down of Fermi Liquid theory occurs because two symmetrically coupled conduction electron channels compete to screen the impurity, which is left partially unscreened, leading to a residual entropy at zero temperature and a NFL groundstate~\cite{emery1992mapping,affleck1993exact,georges1994emery,fabrizio1995crossover,han2022fractional}. 

A rich NFL physics is known to emerge upon enlarging the symmetry of the Kondo problem to a $SU(N)$ spin and $SU(K)$ channels (flavours of conduction electrons) leading to an overscreening of the impurity. In this regime a controlled  theory can be obtained by sending $N,K\rightarrow\infty$ at fixed ratio $\gamma=K/N$. The universal thermodynamic, spectral and transport properties of this multichannel Kondo (MCK) model have been computed and demonstrated emergent conformal invariance and residual entropy~\cite{parcollet1997transition,parcollet1998overscreeneed}, features known to occur in the SYK model. 
However, quantum chaos properties of this class of NFL are not known. Here we fill this gap, specifically by computing the leading low temperature behaviour of the OTOC for the MCK and extract the associated Lyapunov exponent. In the MCK model the impurity spin fractionalizes in auxiliary fermions and bosons which are coupled to conduction electrons. Using the ladder diagrammatics~\cite{gu2019ontherelation} we compute the OTOC of bulk fermions, auxiliary fermions and bosons. We show that all the degrees of freedom have Lyapunov exponent $\lambda_L$ linear in temperature as $T\rightarrow0$, with a prefactor that is tunable with  $\gamma$. For $\gamma=1$, corresponding to a number of channels equal to the size of the spin, the Lyapunov exponent reaches its maximum value, equal to half of the bound value. We develop a diagrammatic technique to compute the OTOC of the physical impurity spin, which is a composite object. Remarkably, we show that the spin OTOC saturates the bound on chaos. The MCK model emerges therefore as the simplest model of maximally chaotic quantum matter, comprising only a strongly coupled quantum spin at the boundary of a non-chaotic system. This result, in addition to the structure of the large $N,K$ theory and its similarity with the SYK model hints towards a gravity analogue for the MCK.

The paper is organised as follows. In the next section \ref{sec:summary} we summarise the main concepts and results of this work, which will be presented in more detail in the rest of the paper. In Sec.~\ref{sec:model} we introduce the multichannel Kondo model and review its large $N$ limit solution leading to Non-Fermi Liquid scaling.  In Sec.~\ref{sec:otoc} we discuss the chaos properties of the model and compute in particular the leading low-temperature behaviour of the OTOC for the fundamental degrees of freedom of the  theory.  In Sec. \ref{sec:spinOTOC} we extend the diagrammatic calculation of the OTOC to the physical spin and obtain the low temperature behaviour of its Lyapunov exponent. In Sec.~\ref{sec:discussion} we discuss our result in the broader perspective of integrability and chaos of quantum impurity models. Finally, Sec. \ref{sec:conclusions} is devoted to conclusions. In the Appendixes we provide technical details on our calculations.

\section{Summary of Main Results}
\label{sec:summary}

In this section we present a summary of the main results of this work, which will be discussed more in detail in the rest of the paper.  The class of models that we consider in this work are quantum impurity models where a single quantum spin is coupled to a bath of non-interacting fermions.
We consider a $SU(N)$ spin coupled to a conduction electron bath featuring $K$ channels of spinful electrons, leading to the multichannel Kondo Model (MCK). 
The Hamiltonian of the model is
\begin{equation}\label{eqn:MCK}
    \hat{H} = \sum_{\Vec{k},i,\alpha} \epsilon_{\Vec{k}} \, \hat{c}^{\dagger}_{\Vec{k} i \alpha } \hat{c}_{\Vec{k} i \alpha } + \frac{J}{N} \sum_{\Vec{k},\Vec{k}^{\prime},i,\alpha,\beta} \hat{S}_{\alpha \beta} \, \hat{c}^{\dagger}_{\Vec{k} i \alpha } \hat{c}_{\Vec{k}^{\prime} i \beta}
\end{equation}
where $\hat{c}^{\dagger}_{\Vec{k} i \alpha} $ creates a conduction electron with moment $\Vec{k}$, channel index $i = 1, \cdots, K$ and SU$(N)$ spin index $\alpha = 1, \cdots, N$. We use a fermionic (antisymmetric) representation for the impurity spin with $N$ auxiliary fermions $\hat{f}_{\alpha}$ with a constraint such that
\begin{equation}\label{eqn:aux_fermions}
    \hat{S}_{\alpha \beta} = \hat{f}_{\alpha}^{\dagger} \hat{f}_{\beta} - \frac{Q}{N} \delta_{\alpha \beta}, \qquad \sum_{\alpha = 1}^{N}  \hat{f}_{\alpha}^{\dagger} \hat{f}_{\alpha} = Q 
\end{equation}
We will consider the limit where both $N$ and $K$ are large keeping the ratio $\gamma = K / N$ fixed. To ensure that the large-$N$ limit is well-defined we also need to take a large spin size $Q \equiv q_{0} N $~\cite{parcollet1998overscreeneed}.  When written in terms of Abrikosov fermions the Kondo coupling describes interaction between impurity fermions and conduction electrons, which can be described by the exchange of a bosonic excitation, leading to a simple diagrammatics in the large $N,K$ limit, that we review in Sec.~\ref{sec:model}.
We are interested in the many-body chaos as encoded in the OTOC of our composite degrees of freedom
\begin{widetext}
\begin{align}
    C_{O}(t_{1},t_{2}) &\sim  \theta(t_{1}) \theta(t_{2}) \sum_{\mu , \nu} \textrm{Tr} \Big[ \sqrt{\rho} \big[ \hat{O}_{\mu}(t_{1}),\hat{O}_{\nu}^{\dagger}(0)  \big]_{\pm}   \sqrt{\rho} \big[ \hat{O}_{\mu}(t_{2}),\hat{O}_{ \nu}^{\dagger}(0)  \big]_{\pm}^{\dagger}  \Big]
\end{align}
\end{widetext}
where $\hat{O}$ can be auxiliary fermions and bosons, conduction electrons as we all as the physical spin, $\mu,\nu$ represent the corresponding degrees of freedom and $[,]_{\pm}$ denotes a commutator or anticommutator whether we consider bosonic or fermionic operators. We are interested in the regime where the OTOC grows exponentially in time, 
$$
C_{O}(t_{1},t_{2})\sim \left(\frac{1}{N}\right)e^{\lambda_L t}
$$
with $t = (t_{1}+ t_{2})/2$ the central time and where we can identify a quantum Lyapunov exponent $\lambda_L$ that we want to compute in the low temperature limit. 

From the point of view of auxiliary fermions  the large $N,K$ theory turns out to be structurally similar to the complex SYK model~\cite{chowdhury2022sachdev} and we show indeed in Sec.~\ref{sec:otoc} that the OTOC of the auxiliary fermions can be computed exactly at low temperature by resumming ladder diagrams. We find that the Lyapunov exponent is linear in temperature with a prefactor that depends on $\gamma=K/N$ and which is maximal at $\gamma=1$. Intriguingly, at this optimal value the Lyapunov exponent is half of maximal bound. 
Similarly we compute the OTOC of the auxiliary bosons and show that the ladder structure gives rise to the same equations as the fermionic auxiliary degrees of freedom. As a result we conclude that the Lyapunov exponent of bosons and fermions is identical and reaches its maximum at $\gamma=1$, where we get $\lambda_L=\pi T$, only half-value away from the bound. Auxiliary fermions and bosons however are not physical (gauge invariant) degrees of freedom in our theory. It is therefore interesting to compute the chaos properties of the physical degrees of freedom.  In Section~\ref{sec:otoc} we compute the OTOC of conduction electrons. We show that the OTOC of the latter can be obtained directly from the OTOC of fermions and bosons and it is also lead to a linear in temperature Lyapunov exponent, with the same prefactor as auxiliary fermions and bosons. This is a remarkable result: conduction electrons in our model are essentially free degrees of freedom in the bulk which interact only with the impurity at their boundary. Such a strong interaction is however sufficient to generate a linear in temperature quantum Lyapunov exponent, at least close to the impurity site. Finally in section~\ref{sec:otoc} we compute the OTOC of the spin, which is composite object in our theory of fermions coupled to bosons. We show that the OTOC of the impurity spin can be still computed in closed form and that, strikingly, its Lyapunov exponent saturates the bound on chaos for $\gamma=K/N=1$. Strong correlations at the boundary of our otherwise non-interacting system lead therefore to maximal quantum chaos. This appears to be equally shared by strongly coupled auxiliary degrees of freedom our quantum spin fractionalises into, a sort of equipartition of many-body quantum chaos. We present a detailed discussion of this remarkable result in Sec.~\ref{sec:discussion} and present our conclusions in Sec.~\ref{sec:conclusions}.

\subsection{Relation to Prior Works}\label{sec:prior}

Here we wish to connect our work to previous literature on many-body chaos in Non-Fermi Liquids and quantum many-body systems. Models of coupled bosons and fermions have been considered before, in the context of Yukawa models for random Dirac Fermions~\cite{kim2021dirac,davis2023quantum}, antiferromagnetic quantum critical points~\cite{lunts2019manybody} or fermions coupled to gauge fields~\cite{patel2017quantum,tikhanovskaya2022maximal}. We note that these works have typically focused only on OTOC of single particle operators. Models of fractionalised degrees of freedom have displayed similar large N theories~\cite{joshi2020deconfined,christos2022critical,christos2022spin}, but their quantum chaos content is not known. In this respect our work goes beyond those studies in that we compute explicitly the OTOC of a fractionalised state of matter. This approach opens up the possibilities for non-perturbative treatments of those couplings in lattice models. 
Recently the information scrambling of the 2CK model has been investigated within the Emery-Kivelson solution~\cite{dora2017information}, leading to an OTOC that reaches a constant at long time. Similarly, the OTOC of a Bose-Fermi Kondo model was evaluated at intermediate temperatures~\cite{han2021quantum}. With respect to these works here we consider the large N,M limit as well as the low temperature limit and extract the leading low-temperature behaviour of the Lyapunov exponent. Similarly, our work differs from the recent concept of boundary chaos introduced in Ref.~\cite{fritzsch2022boundary}, where a quantum circuit with an impurity is studied from the point of view of correlation functions and entanglement.

\section{Model, Large $N,K$ solution and Non-Fermi Liquid behaviour}
\label{sec:model}

\begin{figure}[t!]
   \includegraphics[width=0.5\textwidth]{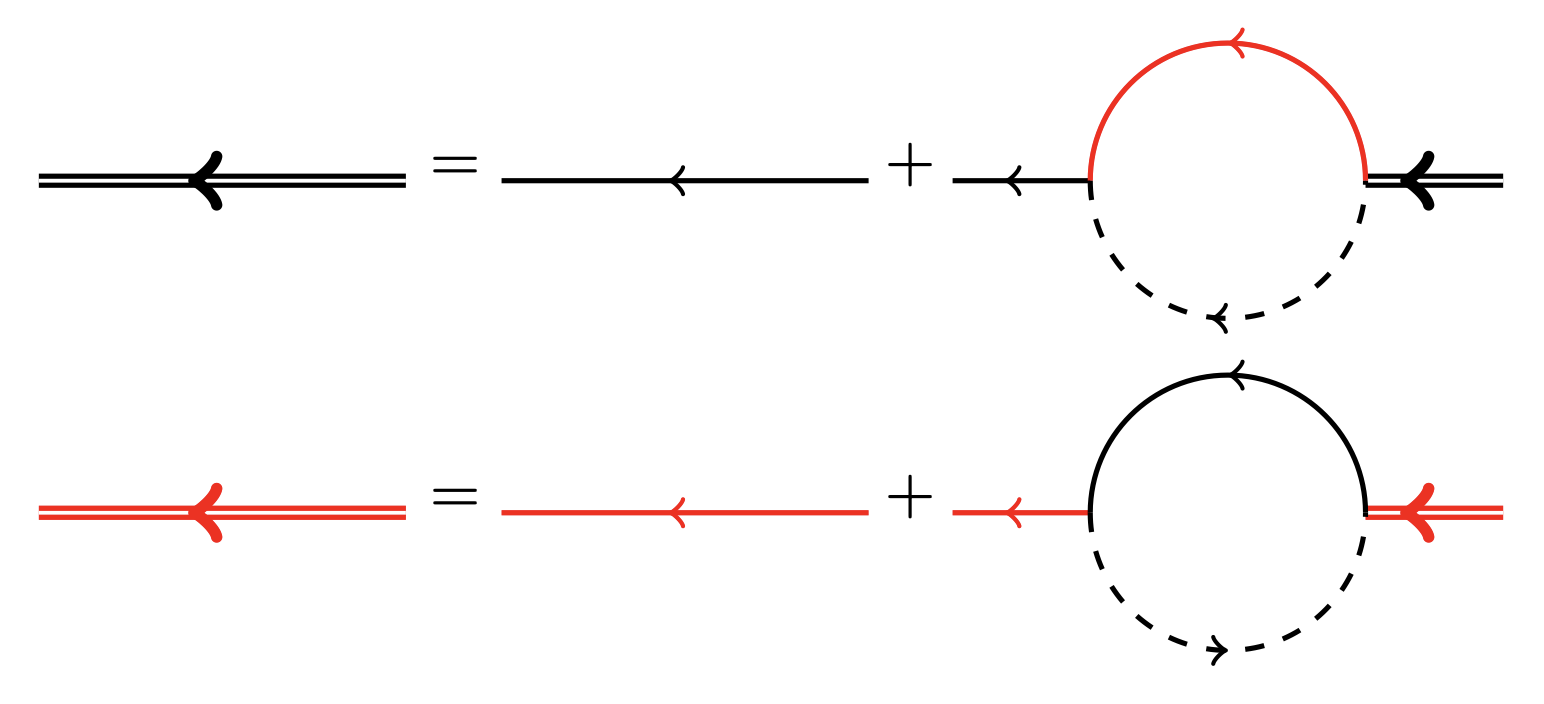}
    \caption{
    \justifying
    \label{fig1}  Feynman diagrams for the Dyson equations (\ref{dyson1}) and (\ref{dyson2}) in the large $N$ limit. Double lines represent dressed propagators and simple lines non-interacting propagators. Black solid lines correspond to the impurity fermion $f$, dashed lines to the conduction electron at the impurity site $c$ and red lines to the auxiliary boson $B$. In these diagrams, each interaction vertex is of order $1 / \sqrt{N}$ and each loop contributes a factor of $N$, so overall the bubbles are of order $\mathcal{O}(1)$ in $N$ and survive in the large $N$ limit. }
\end{figure}

We start reviewing the equilibrium properties of the MCK model, Eq.~(\ref{eqn:MCK}), in the large $N,K$ limit~\cite{parcollet1998overscreeneed}. The partition function for the model, written as a path integral reads
$$
 Z = \int \mathcal{D} [\overline{f},f,\overline{c},c,\overline{B},B,\mu] e^{- S }
 $$ 
where the action $S = S_{0} + S_{int}$ is 
\begin{align}
    S_{0} &= \int_{0}^{\beta} d \tau d \tau^{\prime} \sum_{i \alpha} \overline{c}_{i \alpha }(\tau) (G_{c}^{(0)})^{-1}(\tau,\tau^{\prime}) c_{i \alpha}(\tau^{\prime}) +\nonumber \\
    &+ \int_{0}^{\beta} d \tau \Big\{ \sum_{\alpha} \overline{f}_{\alpha}(\tau) \big[ \partial_{\tau} + i \mu(\tau) \big] f_{\alpha}(\tau) - i q_{0} N \mu(\tau) \Big\} \label{eq:S_{0}} \\
    &+ \frac{1}{J} \int_{0}^{\beta} d \tau \sum_{i} \overline{B}_{i}(\tau) B_{i}(\tau) \nonumber\\
    S_{int} &= - \frac{1}{\sqrt{N}} \int_{0}^{\beta} d \tau \sum_{i \alpha} \Big( \overline{B}_{i}(\tau) \overline{c}_{i \alpha}(\tau) f_{\alpha}(\tau)  \label{eq:S_int}\\
    &+ B_{i}(\tau) \overline{f}_{\alpha}(\tau) c_{i \alpha}(\tau) \Big) \nonumber
\end{align}
In this expression we have introduced auxiliary fermions $f_{\alpha},\overline{f}_{\alpha}$ to represent the spin, as in Eq.~(\ref{eqn:aux_fermions}), 
auxiliary bosons $B_{i},\overline{B}_{i}$ to decouple the Kondo interaction in Eq.~(\ref{eqn:MCK}) and we have denoted with $\mu(\tau)$ the Lagrange multiplier used to impose the constraint on the number of fermions. Furthermore we have integrated out the conduction electrons in the bulk so that $c_{i \alpha} \equiv \sum_{\Vec{k}} c_{\Vec{k} i \alpha } $ is the fermionic field at the impurity site and $G_{c}^{0} = \sum_{\Vec{k}} 1 / (i \omega_{n} - \epsilon_{\Vec{k}}) $ the non-interacting on-site Green's function.  In the large $N$ limit,  the Green's functions
\begin{align}
    G_{f}(\tau,\tau^{\prime}) &= - \frac{1}{N} \sum_{\alpha = 1}^{N} \big\langle f_{\alpha}(\tau) \overline{f}_{\alpha}(\tau^{\prime}) \big\rangle 
    \label{green_f}\\
    G_{B}(\tau,\tau^{\prime}) &= +  \frac{1}{K} \sum_{i = 1}^{K} \big\langle B_{i}(\tau) \overline{B}_{i}(\tau^{\prime}) \big\rangle
    \label{green_B}
\end{align}
satisfy the Schwinger-Dyson equations~\cite{parcollet1998overscreeneed}
\begin{align}
    G_{f}^{-1}(i \omega_{n}) &= i \omega_{n} + \lambda - \Sigma_{f}(i \omega_{n}) \label{dyson1} \\
    G_{B}^{-1}(i \nu_{n}) &= \frac{1}{J} - \Sigma_{B}(i \nu_{n}) \label{dyson2}
\end{align}
with $\Sigma_{f}$ and $\Sigma_{B}$ the self-energies
\begin{equation}
    \Sigma_{f}(\tau) = \gamma \, G_{c}^{0}(\tau) \, G_{B}(\tau)  \qquad
    \Sigma_{B} = G_{c}^{0}(\tau) \, G_{f}(\tau) \label{dyson3}
\end{equation}
Here $\omega_{n} = (2 n + 1 ) \pi / \beta$ and $\nu_{n} = 2 n \pi / \beta$ are respectively the fermionic and bosonic frequencies. Equations (\ref{dyson1}-\ref{dyson3}) can be obtained by the saddle point method after introducing the self-energies as Lagrange multipliers to enforce the definitions of the Green's functions $G_{f}$ and $G_{B}$ and integrating out the fields $f$, $B$ and $c$ to get an effective action written only in terms of the bilocal fields $G$ and $\Sigma$. Here, $\lambda$ is the saddle-point value of the field $\mu(\tau)$ and is fixed by the condition $G_{f}(\tau = 0^{-}) = q_{0}$. Another way of deriving the Schwinger-Dyson equations (\ref{dyson1} - \ref{dyson3}) is to expand the two-point functions (\ref{green_f}),(\ref{green_B}) in powers of the interaction vertex (\ref{eq:S_int}) and keep only the diagrams of order $\mathcal{O}(1)$ in the $1 / N$ expansion. The corresponding Feynamn diagrams are shown in Fig. \ref{fig1}. Notice that if one expands the conduction electron propagator $G_{c}(\tau,\tau^{\prime}) = - \langle c(\tau) \, \overline{c}(\tau^{\prime}) \rangle$, one finds that it is equal to the bare propagator $G_{c}^{0}$ in the large $N$ limit (since in each bubble correction the loop is of order $\mathcal{O}(1)$ and cannot cancel the $1 / N$ factor coming from the two interaction vertices.)  \\
At low energies $\omega, T \ll T_{K}$, where $T_K$ is the Kondo temperature, the Schwinger-Dyson equations (\ref{dyson1} - \ref{dyson3}) have an emergent conformal and U(1) gauge symmetries and one can derive an explicit expression for the Green's functions \cite{parcollet1998overscreeneed,florens2004exact}
\begin{equation}
    G_{f,B}(\tau) = \left\{
    \begin{array}{ll}
        - A_{f,B} \frac{e^{ \alpha \big( \frac{\tau}{\beta} - \frac{1}{2} \big) } }{\cosh{\frac{\alpha}{2}}} \bigg( \frac{\pi}{\beta \sin \big( \frac{\pi \tau}{\beta} \big) }\bigg)^{2 \Delta_{f,B}}  & 0 < \tau < \beta \\
       \xi A_{f,B} \frac{e^{ \alpha \big( \frac{\tau}{\beta} + \frac{1}{2} \big) }}{\cosh{\frac{\alpha}{2}}}  \bigg( \frac{\pi}{\beta \sin \big( - \frac{\pi \tau}{\beta} \big) }\bigg)^{2 \Delta_{f,B} } & - \beta < \tau < 0
    \end{array}
\right. \label{gf_mats}
\end{equation}
$\xi = 1$ (resp. $-1$) for fermions (resp. bosons). To get these expressions, one needs to use the fact that the local non-interacting Green's function of the conduction electron takes the scaling form $ G_{c}^{0}(\tau) = - \rho_{0} \pi / ( \beta \sin ( \pi \tau / \beta ) )  $  (since the corresponding density of states is independent of temperature). The scaling dimensions of the fermions and bosons are given by
\begin{equation}
    2 \Delta_{f} = \frac{1}{1 + \gamma} \qquad 2 \Delta_{B} = \frac{\gamma}{1 + \gamma}
\end{equation}
$\alpha $ is an asymmetry paramater which is linked to the spin representation parameter $q_{0}$ by
\begin{equation}
    \alpha = \ln \frac{ \sin \Big( \frac{\pi q_{0}}{ 1 + \gamma}  \Big) }{ \sin \Big( \frac{\pi (1 - q_{0})}{1 + \gamma} \Big)  }
\end{equation}
Finally, the product of amplitudes $A_{f} A_{B}$ satisfies the constraint
\begin{equation}
    2 \Delta_{B} = - 2 \gamma A_{f} A_{B} \rho_{0} \Gamma \big( 1 - 2 \Delta_{B} \big) \Gamma \big( 2 \Delta_{B} \big) \frac{ \Big| \sin \Big( \pi \Delta_{B} - i \frac{\alpha}{2} \Big) \Big|^{2} }{ \cosh^{2}\Big( \frac{\alpha}{2} \Big) }
\end{equation}
Details about the origin of these formulas can be found in \cite{parcollet1998overscreeneed}. The saddle point large $N,M$ solution of the MCK model describes therefore a Non-Fermi Liquid associated with the overscreening of the impurity spin by the $K$ channels of the bath. In the next section we will compute the OTOC properties of this model.

\section{Out of Time Order Correlator and Lyapunov Exponent}\label{sec:otoc}

In this section we compute the OTOC for the different degrees of freedom entering our theory, namely auxiliary fermions and bosons as well as conduction electrons. In particular we will derive Bethe-Salpeter equations for these objects and extract the leading low-temperature behaviour of the Lyapunov exponent $\lambda_{L}$, or scrambling rate, describing the short-time exponential growth of the OTOC.
\subsection{f-fermion and B-boson OTOC}

\begin{figure}[t!]
    \centering
    \includegraphics[scale = 0.2]{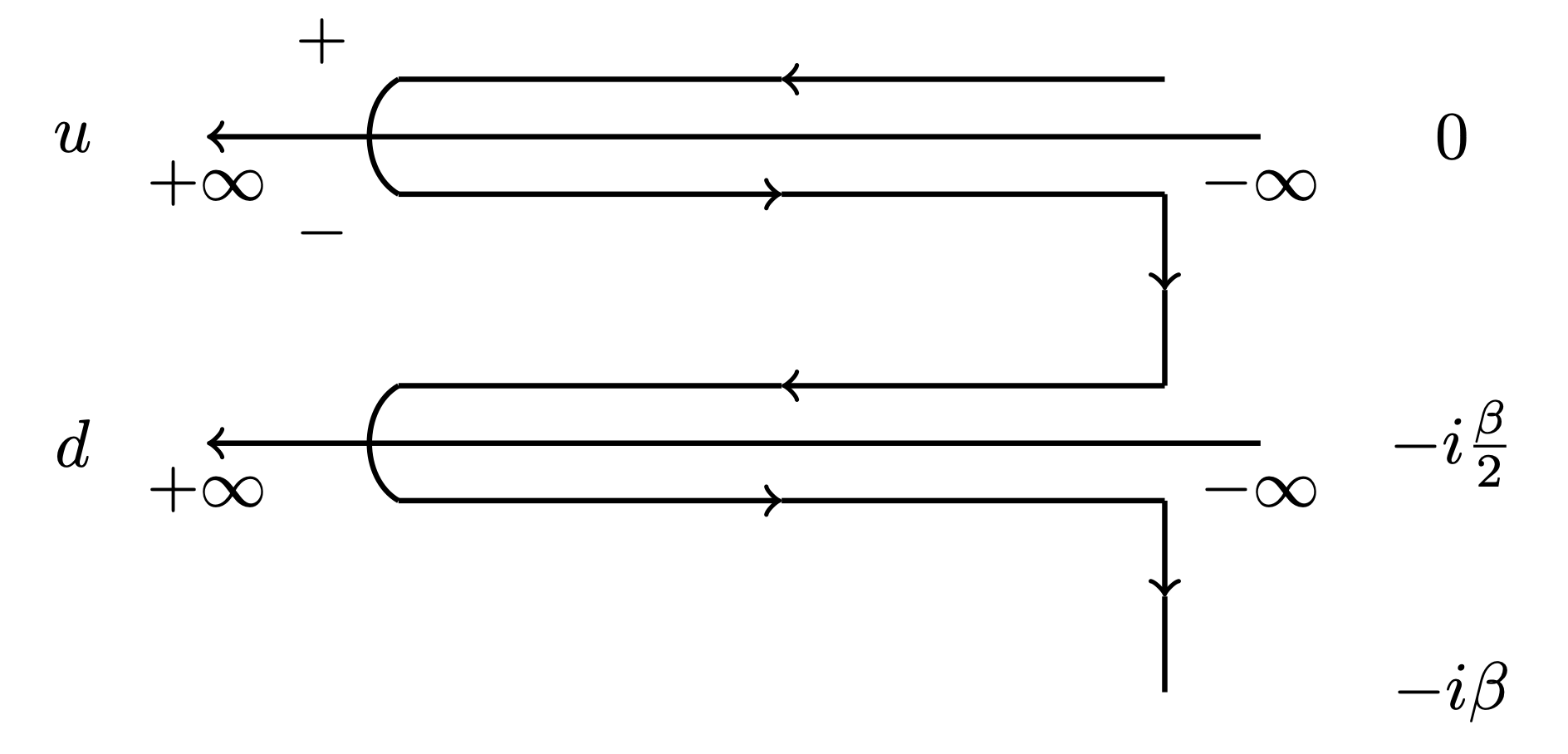}
\caption{
    \justifying
    \label{fig2}  Two-fold Keldysh contour used to define the regularised OTOC that we consider in this work. The imaginary time evolution is broken down into two time-intervals, each with an evolution for an interval $-i\beta/2$ and the real-time evolution can occur on the up ($u$) or down ($d$) double contour.}
    \label{fig2}
\end{figure}

\begin{figure}[t!]
    \centering
    \includegraphics[scale = 0.25]{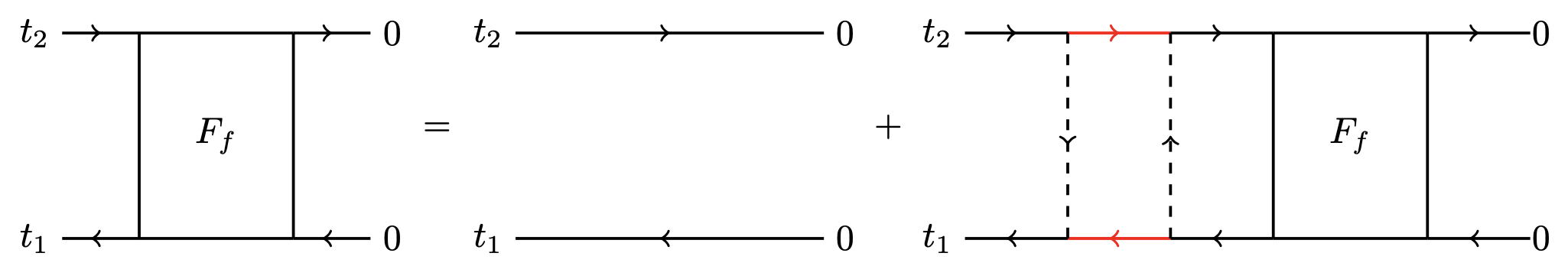}
   \caption{
    \justifying
    \label{fig3}  Resummation of ladder diagrams for the OTOC of the auxiliary fermions, leading to a Bethe Salpether equation for $F_f(t_1,t_2)$. Dashed lines are conduction electrons propagators while full red-lines are bosonic ones. The resulting expression is given in Eq.~(\ref{eqn:BS_fermion}).}
\end{figure}

We start with the regularized retarded OTOC of the auxiliary fermions
\begin{equation}
\begin{split}
    C_{f}(t_{1},t_{2}) &= \frac{\theta(t_{1}) \theta(t_{2})}{N^{2} } \sum_{\alpha , \beta = 1}^{N} \textrm{Tr} \Big[ \sqrt{\rho} \big\{ \hat{f}_{\alpha}(t_{1}),\hat{f}_{\beta}^{\dagger}(0)  \big\} \\
    &\times \sqrt{\rho} \big\{ \hat{f}_{\alpha}(t_{2}),\hat{f}_{ \beta}^{\dagger}(0)  \big\}^{\dagger}  \Big] \, .
\end{split}
\end{equation}
This correlator can be computed using the two-fold Keldysh contour shown on Fig. \ref{fig2}. This complex time contour consists of two real time folds separated by an imaginary time segment of length $- i \beta /2$. The position of the fields on the contour is specified by a fold index $\sigma = u,d$ and a branch index $s = +,-$ for forward and backward time evolution respectively. To proceed it is convenient to perform a Keldysh rotation on each fold and define the classical and quantum components
\begin{align}
    f_{\alpha}^{\sigma, cl/q}(t) &= \frac{ f_{\alpha}^{\sigma, +}(t) \pm f_{\alpha}^{\sigma, -}(t) }{\sqrt{2}} \, ,   \\
    \overline{f}_{\alpha}^{\sigma, cl/q}(t) &= \frac{ \overline{f}_{\alpha}^{\sigma, +}(t) \mp \overline{f}_{\alpha}^{\sigma, -}(t) }{\sqrt{2}}  \, .
\end{align}
A similar transformation is done for the auxiliary bosons and the conduction electrons. Within the two-fold Keldysh formalism, the regularized retarded OTOC can be rewritten as the four-point function (see Appendix~\ref{app:Keldysh_double})
\begin{equation}
    C_{f}(t_{1},t_{2}) = \frac{1}{N^{2} } \sum_{\alpha , \beta = 1}^{N} \big\langle f_{ \alpha}^{d,cl}(t_{1}) \overline{f}_{ \beta}^{d,cl}(0) \overline{f}_{ \alpha}^{u,q}(t_{2}) f_{ \beta}^{u,q}(0) \big\rangle \, ,
\end{equation}
where $\langle \cdots \rangle $ denotes the average over the two-fold Keldysh contour. We now expand this correlator in powers of the interaction vertex and keep only the diagrams at leading order $1 / N$. For convenience, we define $ C_{f}(t_{1},t_{2}) \equiv F_{f}(t_{1},t_{2}) / N$. The $1/N$ diagrams form and infinite series of ladder diagrams, which after resummation are represented in Fig. \ref{fig3}. In these Feynman diagrams, horizontal lines represent retarded/advanced Green's functions $G^{R/A}$ and vertical lines are Wightman Green's functions $G^{K}_{u d}$ or $G^{K}_{d u}$ (see Appendix~\ref{app:realtimeGF}). The resummation of all the ladder diagrams shown in Fig. \ref{fig3} corresponds to the Bethe-Salpeter (BS) equation
\begin{equation}\label{eqn:BS_fermion}
\begin{split}
    F_{f}(t_{1},t_{2}) &= F_{f}^{(0)}(t_{1},t_{2}) + \frac{\gamma}{4} \int d t_{3} \, d t_{4} \, K_{f}(t_{1},t_{2},t_{3},t_{4}) \\
    &\times F_{f}(t_{3},t_{4})
\end{split}
\end{equation}
where $ F_{f}^{(0)} (t_{1},t_{2}) = G_{f}^{R}(t_{1}) G_{f}^{A}(-t_{2}) $ and  the kernel $K_{f}$ is given by
\begin{align}
    K_{f}(t_{1},t_{2},t_{3},t_{4}) &= \int d t \, d t^{\prime} K^{(1)}_{f}(t_{1},t_{2},t,t^{\prime}) \nonumber \\
    &\times K^{(2)}_{f}(t,t^{\prime},t_{3},t_{4})
\\
   K^{(1)}_{f}(t_{1},t_{2},t_{3},t_{4}) &= G_{f}^{R}(t_{1},t_{3}) \, G_{f}^{A}(t_{4},t_{2}) \, G_{c,d u}^{K}(t_{3},t_{4}) \\
    K^{(2)}_{f}(t_{1},t_{2},t_{3},t_{4}) &= G_{B}^{R}(t_{1},t_{3}) \, G_{B}^{A}(t_{4},t_{2}) \, G_{c,u d}^{K}(t_{4},t_{3})
\end{align}
Explicit expressions for the Green's functions entering the BS equation are given in Appendix~\ref{app:realtimeGF}.
With respect to the standard ladder diagrammatics of the OTOC for the SYK model~\cite{kitaev2018softmode} we note that here the kernel $K_f$ is itself expressed as a convolution over product of fermionic, bosonic and conduction electrons Green's function. This is a direct consequence of the structure of the Feynman vertex of the theory. Similar structure appears in other theories with fermions coupled to bosons~\cite{lunts2019manybody,kim2020scrambling,kim2021dirac,tikhanovskaya2022maximal}. We note that a similar equation was written for the Bose-Fermi Kondo model~\cite{han2021quantum} and solved numerically to extract the Lyapunov exponent at high temperature. Here instead we derive the scrambling rate $\lambda_{L}$ in the $T \rightarrow 0$ limit, by making use of the conformal expressions of the Green's functions given in Appendix~\ref{app:realtimeGF}. Since we expect $F_{f}$ to grow exponentially, we neglect the first term $F_{f}^{(0)}$ in the ladder equation. The Bethe-Salpeter equation becomes 
\begin{equation}
\begin{split}
    F(t_{1}&,t_{2}) = C e^{ i \frac{\alpha}{\beta} t_{1 2} } \int d t_{3} \, d t_{4} \, d t_{5} \, d t_{6} \Tilde{G}_{f}^{R}(t_{1 3}) \Tilde{G}_{f}^{R}(t_{2 4})  \\
    &\times \Tilde{G}_{c}^{W}(t_{3 4}) \Tilde{G}_{B}^{R}(t_{3 5}) \Tilde{G}_{B}^{R}(t_{ 4 6 }) \Tilde{G}_{c}^{W}(t_{ 5 6 })   e^{ - i \frac{\alpha}{\beta} t_{5 6} } F(t_{5},t_{6}) \, ,
\end{split}
\end{equation}
where we have used the notation $t_{ij} \equiv t_{i} - t_{j}$ and with $\Tilde{G}_{f,B}^{R}(t)$ and $\Tilde{G}_{c}^{W}(t)$ defined by
\begin{equation}
    \Tilde{G}_{f,B}^{R}(t) = \theta(t) \bigg( \frac{\pi}{ \beta \sinh \big( \frac{\pi t}{\beta} \big) } \bigg)^{2 \Delta_{f, B}},  \Tilde{G}_{c}^{W}(t) = \frac{\pi}{\beta \cosh \big( \frac{\pi t}{\beta} \big) } \, ,
    \label{eq:MCK_GF_real_time}
\end{equation}
and with
\begin{equation}
    C = 16 \frac{ \Delta_{f} \Delta_{B} }{ \Gamma ( 1 - 2 \Delta_{f} )^{2} \Gamma ( 1 - 2 \Delta_{B} )^{2}  } \, .
\end{equation}
To get rid of the exponential phase factors we define $ \Tilde{F}(t_{1},t_{2}) = e^{ - i \frac{\alpha}{\beta} t_{1 2} } F(t_{1},t_{2})$. The BS equation takes the form of an eigenvalue problem $ \Tilde{K} \cdot \Tilde{F} = k \, \Tilde{F} $ with $ k =1$. To find the Lyapunov exponent, we assume that $\Tilde{F}$ has an exponential growth

\begin{equation}
    \Tilde{F}(t_{1},t_{2}) = e^{ \lambda_{L} \frac{t_{1} + t_{2 }}{2}   } \Tilde{f}(t_{1} - t_{2}) \, ,
    \label{eq:expo_ansatz}
\end{equation}
and we determine the eigenvalues and eigenvectors of the kernel $\Tilde{K}$ in this subspace. $\lambda_{L}$ should be seen as a free parameter, which is fixed at the end of the computation by setting $k(\lambda_{L}) = 1$. Plugging the exponential ansatz in the BS equation and taking the Fourier transform with respect to $t_{1} - t_{2}$ we obtain
\begin{equation}
\begin{split}
    k(\lambda_{L} ) \, \Tilde{f}(\omega) &= C \, \Tilde{G}_{f}^{R} \Big( \omega + i \frac{\lambda_{L}}{2} \Big) \Tilde{G}_{f}^{R} \Big( - \omega + i \frac{\lambda_{L}}{2} \Big) \\
    &\times \int \frac{d \omega^{\prime} }{2 \pi} \Tilde{K}(\omega, \omega^{\prime}) \Tilde{f}(\omega^{\prime})
\end{split}
    \label{eq:BS_freq}
\end{equation}
with
\begin{equation}
\begin{split}
    \Tilde{K}(\omega,\omega^{\prime}) &= \int \frac{d \Omega}{2 \pi} \Tilde{G}_{B}^{R} \Big( \Omega + i \frac{\lambda_{L}}{2} \Big) \Tilde{G}_{B}^{R} \Big( - \Omega + i \frac{\lambda_{L}}{2} \Big) \\
    &\times \Tilde{G}_{c}^{W}(\omega - \Omega) \Tilde{G}_{c}^{W}(\Omega - \omega^{\prime})
\end{split}
\end{equation}
Using the Fourier transforms of (\ref{eq:MCK_GF_real_time}), we can eventually rewrite the BS equation like
\begin{equation}
\begin{split}
k(h) \Tilde{f} (u) &= 16  \frac{ \Delta_{f} \Delta_{B}}{ (2 \pi )^{2} } \, \bigg| \frac{ \Gamma \big( \Delta_{f} + \frac{h}{2} - i u \big)  }{ \Gamma \big(1 - \Delta_{f} + \frac{h}{2} - i u \big)  }  \bigg|^{2} \\ &\times  \int d u^{\prime} \, d v \left[ \bigg| \frac{ \Gamma \big( \Delta_{B} + \frac{h}{2} - i v \big)  }{ \Gamma \big(1 - \Delta_{B} + \frac{h}{2} - i v \big)  }  \bigg|^{2} \right. \\
&\times \left. \Big| \Gamma \Big( \frac{1}{2} + i (u - v) \Big) \Big|^{2} \Big| \Gamma \Big( \frac{1}{2} + i (u^{\prime} - v) \Big) \Big|^{2}  \Tilde{f}(u^{\prime}) \right]
\end{split}
\end{equation}
Here we have defined  $h = \lambda_{L} / (2 \pi T) $ and made the change of variable $u = \omega / (2 \pi T)$. Using the identity on gamma functions
\begin{equation}
\begin{split}
    \int_{- \infty}^{+ \infty} d u^{\prime} \, \Big| &\Gamma \Big( \frac{1}{2} + i (u - u^{\prime}) \Big) \Big|^{2} \, \Big| \Gamma \Big( \frac{a}{2} + i u^{\prime}  \Big) \Big|^{2} \\
    &= \frac{2 \pi}{a} \, \Big| \Gamma \Big( \frac{1}{2} + \frac{a}{2} + i u  \Big) \Big|^{2}
    \end{split}
\end{equation}
we see that the function $\Tilde{f}(u)$
\begin{equation}
    \Tilde{f}(u) = \Big| \Gamma \Big( \Delta_{f} + \frac{h}{2} + i u \Big) \Big|^{2}
    \label{eq:eigenvec}
\end{equation}
is solution of the eigenproblem with eigenvalue
\begin{equation}
    k(h) = 16 \, \frac{ \Delta_{f} \Delta_{B} }{ ( 2 \Delta_{f} + h ) (2 \Delta_{B} + h )} \label{eq:eigen}
\end{equation}
Setting $k(h) = 1$ and keeping only the positive solution we finally obtain
\begin{equation}\label{eqn:lyapunov}
    \lambda_{L} =  \varphi(\gamma) \, 2 \pi T, \qquad \varphi(\gamma) = \frac{- 1 + \sqrt{1 + 12 \, \frac{\gamma}{(1 + \gamma )^{2}}} }{2} 
\end{equation}
\begin{figure}[t!]
    \centering
    \includegraphics[scale = 0.55]{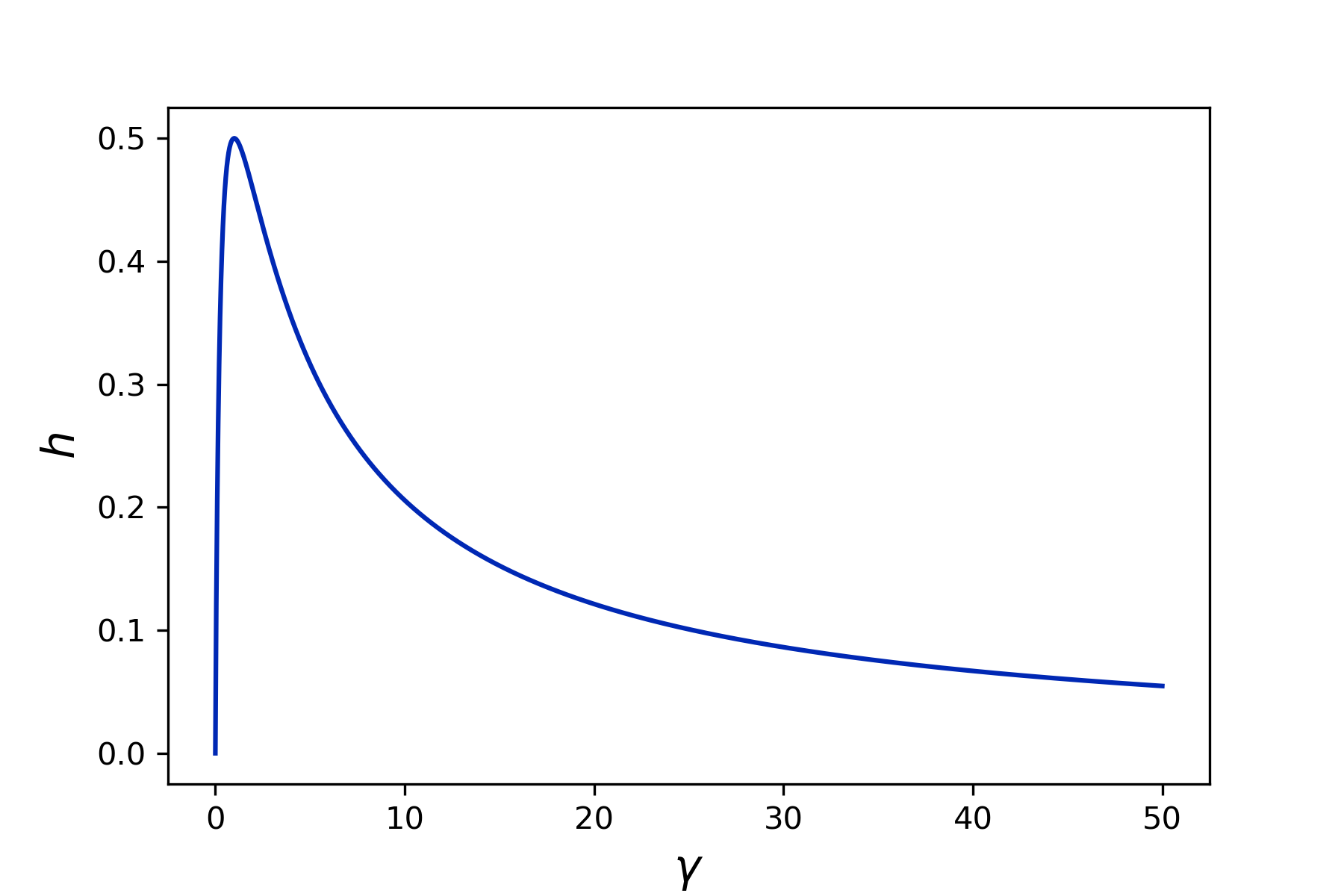}
    \caption{
    \justifying     \label{fig4} 
    Normalised Lyapunov exponent $h = \lambda_{L} / (2 \pi T )$ of auxiliary fermions and bosons in the MCK model, as a function of $\gamma=K/N$, the ratio between number of channels and spin indexes. We see that the Lyapunov exponent vanishes for both small and large $\gamma$ and has a maximum for $\gamma=1$. The maximal value obtained is $h=0.5$, corresponding to $\lambda_L=\pi T$.}
\end{figure}
We find that the scrambling rate is linear in temperature, with a prefactor which depends on the ratio $\gamma$. The function $\varphi(\gamma)$ is plotted in Fig. \ref{fig4}. It vanishes for $\gamma = 0$ and $\gamma = \infty$ which corresponds to the scaling dimensions of the fields $(\Delta_{f},\Delta_{B}) = (\frac{1}{2},0)$ and $(\Delta_{f},\Delta_{B}) = (0,\frac{1}{2})$ respectively. It is maximal when $\gamma = 1$ for which $\Delta_{f} = \Delta_{B} = \frac{1}{4}$ and reach the value $\varphi(1) = \frac{1}{2}$. Notice that at the maximum $\gamma = 1$ both fields $f$ and $B$ have the same scaling dimension as fermions described by the SYK Hamiltonian. However, differently from the SYK case, here the Lyapunov exponent does not saturate the chaos bound but only reaches a value that is exactly half of the maximal chaos, $\lambda_L=\lambda_L^{\rm max}/2=\pi T$. We now turn to the computation of the OTOC of the auxiliary bosons $B$.
As done previously, we define the regularized retarded OTOC
\begin{equation}
\begin{split}
    C_{B}(t_{1},t_{2}) &= \frac{\theta(t_{1}) \theta(t_{2})}{K^{2} } \sum_{i,j = 1}^{K} \textrm{Tr} \Big[ \sqrt{\rho} \big[ \hat{B}_{i}(t_{1}),\hat{B}_{j}^{\dagger}(0)  \big] \\
    &\times \sqrt{\rho} \big[ \hat{B}_{i}(t_{2}),\hat{B}_{ j}^{\dagger}(0)  \big]^{\dagger}  \Big] \, .
\end{split}
\end{equation}
The OTOC can be rewritten as a four-point function on the on the two-fold Keldysh contour introduced above
\begin{equation}
C_{B}(t_{1},t_{2}) = \frac{-1}{K^{2} } \sum_{i,j = 1}^{K} \big\langle B_{ i}^{d,cl}(t_{1}) \overline{B}_{ j}^{d,q}(0) \overline{B}_{i}^{u,cl}(t_{2}) B_{j}^{u,q}(0) \big\rangle 
\end{equation}
Doing a similar diagrammatic expansion as for the auxiliary $f$-fermions, we find that the $\frac{1}{N}$ part $F_{B}$ of $C_{B}$ satisfies the following ladder equation in the limit $N \rightarrow \infty$
\begin{equation}\label{eqn:BS_boson}
\begin{split}
    F_{B}(t_{1},t_{2}) &= F^{(0)}_{B}(t_{1},t_{2}) + \frac{\gamma}{4} \int d t_{3} \, d t_{4} K_{B}(t_{1},t_{2},t_{3},t_{4}) \\
    &\times F_{B}(t_{3},t_{4}) 
\end{split}
\end{equation}
where
\begin{align}
    K_{B}(t_{1},t_{2},t_{3},t_{4}) &= \int d t \, d t^{\prime} K_{B}^{(1)}(t_{1},t_{2},t,t^{\prime}) \\
    &\times K_{B}^{(2)}(t,t^{\prime},t_{3},t_{4})
\\
   K_{B}^{(1)}(t_{1},t_{2},t_{3},t_{4}) &= G_{B}^{R}(t_{1},t_{3}) G_{B}^{A}(t_{4},t_{2}) G_{c,u d}^{K}(t_{4},t_{3}) \\
    K_{B}^{(2)}(t_{1},t_{2},t_{3},t_{4}) &= G_{f}^{R}(t_{1},t_{3}) G_{f}^{A}(t_{4},t_{2}) G_{c,d u}^{K}(t_{3},t_{4})
\end{align}

\begin{figure}[t!]
    \centering
    \includegraphics[scale = 0.2]{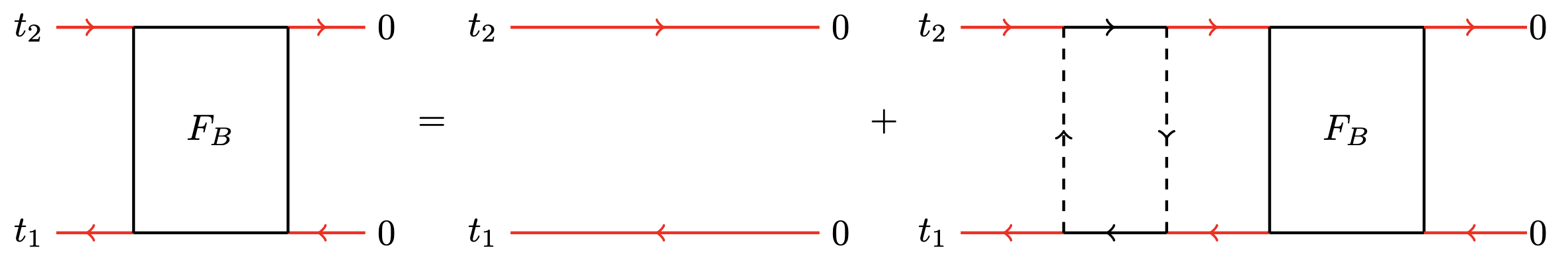}
    \caption{
    \justifying
    \label{fig5}  Resummation of ladder diagrams for the OTOC of the auxiliary bosons, leading to a Bethe Salpether equation for $F_B(t_1,t_2)$. Dashed lines are conduction electrons propagators while full red-lines are bosonic ones. The resulting expression is given in Eq.~(\ref{eqn:BS_boson}).}
\end{figure}

The corresponding Feynman diagrams are shown in Fig. \ref{fig5}. We note that the kernel $K_B$ entering the bosonic OTOC has a similar structure with respect to the fermionic one, namely two loops made with fermionic (bosonic) Green's function and vertical (Wightman) conduction electons one. In the low-temperature limit, using the conformal expressions for the Green's function, we conclude that the two kernels 
 $K_{f}$ and $K_{B}$ are equal upon changing the scaling dimensions of auxiliary fermions and bosons, i.e. upon sending $\Delta_{f} \leftrightarrow \Delta_{B}$. Then the computation of the Lyapunov exponent proceeds in exactly the same way as above. The eigenvector (\ref{eq:eigenvec}) has the same form with $\Delta_{f}$ replaced by $\Delta_{B}$. Remarkably, the eigenvalue $k(h)$ is symmetric under exchange of $\Delta_{f}$ and $\Delta_B$. As a result the scrambling rate for the auxiliary bosons $B$ is the same as for impurity fermions $f$ and given again by Eq.~(\ref{eqn:lyapunov}) and plotted in Fig.~\ref{fig4}. To summarize in our MCK the physical spin fractionalizes in strongly coupled auxiliary fermions and bosons. Their scrambling properties at low temperature turns out to be identical, a Lyapunov exponent $\lambda_L$ linear in temperature with a prefactor that depends on $\gamma$. Intriguingly for $\gamma=K/N=1$ both modes acquire their maximal scrambling rate which is however only half of the maximal value allowed by the bound on chaos. This result is therefore different from other models of fermions coupled to bosons in which the two sectors acquire a Lyapunov exponent parametrically smaller one from the other, as well as from the maximum allowed value~\cite{lunts2019manybody} 
 
 We should remark at this stage that these auxiliary degrees of freedom are not physical, gauge invariant objects but only serve to represent the quantum spin. In order to fully elucidate the many-body chaos content of our MCK model it is therefore crucial to evaluate the OTOC of physical degrees of freedom, such as conduction electrons and the impurity spin as we will do in the next sections.
 
\begin{figure*}[t!]
\centering
    \includegraphics[scale = 0.25]{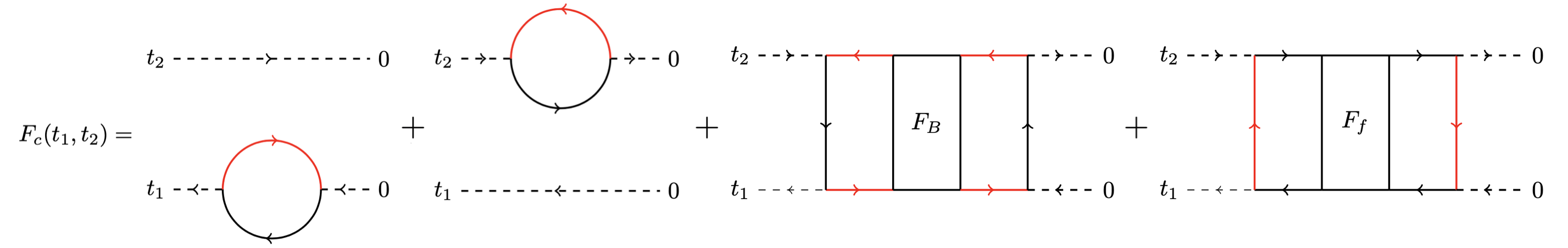}
    \caption{
    \justifying     \label{fig6} 
   Resummation of ladder diagrams for the OTOC of the conduction electrons, $ F_{c}(t_{1},t_{2})$. Dashed lines are conduction electrons propagators while full red-lines are bosonic ones. The resulting expression is given in Eq.~(\ref{eqn:otoc_celectrons1}-\ref{eq:F_f_c}).}
\end{figure*}

\subsection{Conduction electrons}

We now turn to the computation of the OTOC of the conduction electrons at the impurity site
\begin{equation}
\begin{split}
    C_{c}(t_{1},t_{2}) &= \frac{\theta(t_{1}) \theta(t_{2})}{N^{2} K^{2} } \sum_{i, j = 1}^{K} \sum_{\alpha , \beta = 1}^{N} \textrm{Tr} \Big[ \sqrt{\rho} \big\{ \hat{c}_{i \alpha}(t_{1}),\hat{c}_{j \beta}^{\dagger}(0)  \big\} \\
    &\times \sqrt{\rho} \big\{ \hat{c}_{i \alpha}(t_{2}),\hat{c}_{j \beta}^{\dagger}(0)  \big\}^{\dagger}  \Big]
\end{split}
\end{equation}
which can be represented as the four-point function on the two-fold Keldysh contour like
\begin{equation}
\begin{split}
C_{c}(t_{1},t_{2}) &= \frac{1}{N^{2} K^{2} } \sum_{i,j = 1}^{K} \sum_{\alpha , \beta = 1}^{N} \Big\langle c_{ i \alpha}^{d,cl}(t_{1}) \overline{c}_{ j \beta}^{d,cl}(0)  \\
&\times \overline{c}_{i \alpha}^{u,q}(t_{2}) c_{j \beta}^{u,q}(0) \Big\rangle
\end{split}
\end{equation}
We expand this correlator in powers of the interaction vertex including $1/N$ corrections
\begin{equation}
    C_{c}(t_{1},t_{2}) =\frac{1}{N K} G^{R,0}_{c}(t_{1}) G_{c}^{A,0}(-t_{2}) + \frac{1}{N} \frac{1}{N K} F_{c}(t_{1},t_{2}) + \cdots
\end{equation}
In the large $N$ limit, $F_{c}(t_{1},t_{2})$ satisfies the ladder equation shown on Fig. \ref{fig6}.
\begin{equation}\label{eqn:otoc_celectrons1}
    F_{c}(t_{1},t_{2}) = F_{c}^{(0)}(t_{1},t_{2}) + F^{(B)}_{c}(t_{1},t_{2}) +  F^{(f)}_{c}(t_{1},t_{2}) 
\end{equation}

The first two diagrams, which correspond to $ F_{c}^{(0)} $, are the $1 / N$ bubble corrections to the bare electronic propagator. The exponential growth will come from the last two diagrams $F_{c}^{(B)}$ and $F_{c}^{(f)}$ and we will focus on these two. Notice that this ladder equation is not a self-consistent equation as the previous BS equations for fermions and bosons, i.e. $F_{c}$ only appears on the left hand side at the leading order. This is ultimately due to the structure of the diagrammatics in our MCK model, with a vertex coupling fermions, bosons and conduction electrons.
As such  we can evaluate the OTOC of conduction electrons directly since we already know $F_{f}$ and $F_{B}$ from the previous section. 

We detail the derivation for the diagram involving $F_{f}$; it is then straightforward to adapt the computation to the other one. The fermionic contribution $ F^{(f)}_{c}(t_{1},t_{2})$ can be read off from Fig.\ref{fig6} and it is given by
\begin{equation}
\begin{split}
F^{(f)}_{c}(t_{1},t_{2})&=\int \prod_{i=3}^{i=6}dt_i
G_{c}^{R}(t_{1},t_{3}) G_{c}^{A}(t_{4},t_{2}) G_{B,ud}^{K}(t_{4},t_{3}) \\
&\times F_{f}(t_{3},t_{4},t_{5},t_{6})  G_{c}^{R}(t_{5}) G_{c}^{A}(-t_{6}) G_{B,d u}^{K}(t_{5},t_{6})
\label{eq:F_f_c}
\end{split}
\end{equation}
In this expression, $F_{f}(t_{1},t_{2},t_{3},t_{4})$ is a generalisation of $F_{f}(t_{1},t_{2})$ where we replace the times $0$ by $t_{3}$ and $t_{4}$ on the right end of the ladders in (\ref{fig4}). To compute this quantity, we replace the exponential ansatz (\ref{eq:expo_ansatz}) by $\Tilde{F}(t_{1},t_{2},t_{3},t_{4}) = e^{\lambda_{L} \frac{t_{1} + t_{2} - t_{3} - t_{4}}{2} } f_{1}(t_{1} - t_{2}) f_{2}(t_{3} - t_{4}) $. After Fourier transform with respect to $t_{1} - t_{2}$ and $t_{3} - t_{4}$, $f_{2}$ drops out of the ladder equation and $f_{1}$ satisfies (\ref{eq:BS_freq}) and so is equal to (\ref{eq:eigenvec}). Taking the inverse Fourier transform we obtain
\begin{align}
    F_{f}(t_{1},t_{2},t_{3},t_{4}) = e^{ i \frac{\alpha}{\beta} (t_{1} - t_{2} ) } e^{ \lambda_{L} \frac{t_{1} + t_{2}- t_{3} - t_{4}  }{2} }\times\nonumber\\
     \times \frac{ \Gamma \big( \Delta_{f} + \frac{h}{2} \big) }{(2 \pi T )^{2 \Delta_{f} + h - 1 } } \bigg(  \frac{\pi}{\beta \cosh \big( \frac{\pi t_{1 2} }{\beta} \big) }   \bigg)^{2 \Delta_{f} + h} f_{2}(t_{3} - t_{4})
\end{align}
Pluggin this expression in (\ref{eq:F_f_c}), the integral over $(t_{5},t_{6})$ factorises and gives a constant. If we replace $G_{B,ud}^{K}$ by its conformal expression given in Appendix~\ref{app:realtimeGF}, we finally obtain
\begin{align}
F^{(f)}_{c}(t_{1},t_{2})= \textrm{cte} \times e^{ \lambda_{L} \frac{t_{1} + t_{2}}{2}} H(t_{1} - t_{2})
\end{align}
with
\begin{equation}
\begin{split}
    H(t) &= \int \frac{d \omega}{2 \pi} e^{- i \omega (t_{1} - t_{2}) } G_{c}^{R,0} \Big( \omega + i \frac{\lambda_{L}}{2} \Big) \\
    &\times G_{c}^{A,0} \Big( - \omega + i \frac{\lambda_{L}}{2} \Big) f_{c}(\omega),
    \end{split}
\end{equation}
with $ f_{c}(\omega) \equiv  \bigg(  \frac{\pi}{\beta \cosh \big( \frac{\pi t }{\beta} \big) }   \bigg)^{1+ h} $.
 One would find a similar expression for $F_{c}^{(B)}$ and so we can conclude that $F_{c}$ grows exponentially $F_{c}(t_{1},t_{2}) \sim e^{\lambda_{L} \frac{t_{1} + t_{2}}{2} }$ witht the same Lyapunov exponent $\lambda_{L}$ as the impurity fermions and the auxiliary bosons. This result is particularly remarkable, since the conduction electrons are essentially non-interacting degrees of freedom in their bulk and hence non-chaotic (their Lyapunov exponent far away from the impurity should vanish). However due to the strong coupling at the boundary with the impurity spin they develop many-body chaos with a linear-in temperature Lyapunov exponent.  This suggests that the OTOC of the conduction electrons should develop a non-trivial spatial dependence, analogue to the Kondo screening cloud. We leave the investigation of this interesting feature to future work. Next we focus on the OTOC and scrambling properties of the impurity spin.

\begin{figure*}[t!]
    \centering
    \includegraphics[scale = 0.3]{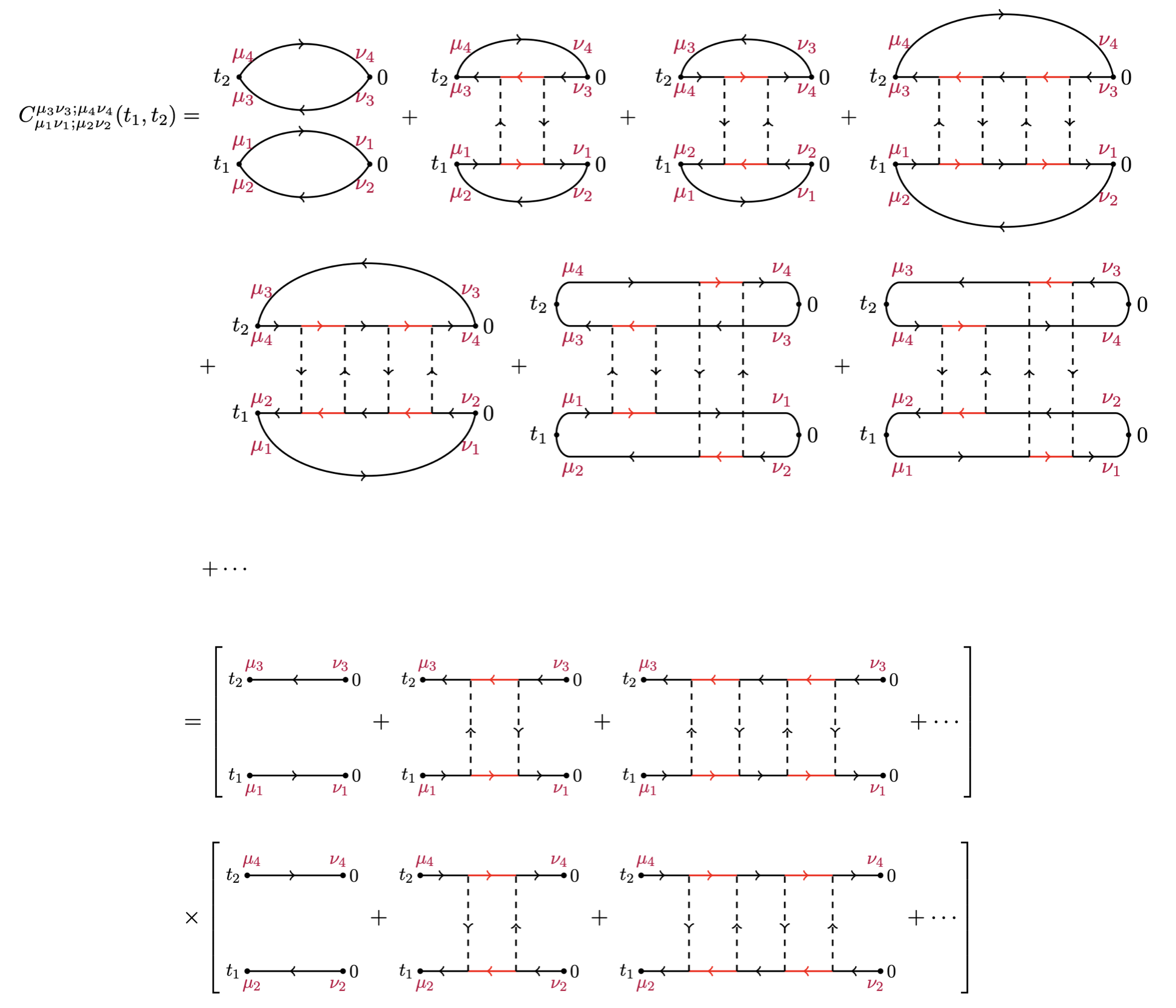}
    \caption{
    \justifying
    \label{fig7}  Feynman diagrams for the impurity spin OTOC. Resummation of ladder diagrams for the OTOC of the impurity spin fermions. Dashed lines are conduction electrons propagators while full red-lines are bosonic ones. The resulting expression is given in Eq.~\ref{eqn:wick_OTOC_Spin}.}
\end{figure*}

\section{Impurity spin OTOC}\label{sec:spinOTOC}

In the previous sections we have computed the OTOC of auxiliary fermions, bosons and conduction electrons. These types of correlators can be written as two-particle Green's functions in the doubled Keldysh contour and their evaluation can be done, to large extent, following similar calculations for the SYK model. In this section we tackle a more challenging OTOC correlator of the impurity spin, which as discussed, is a composite object in our theory of the MCK model. As we are going to show this results in a different diagrammatic structure, summing simple ladder diagrams will not be enough. Remarkably we will be able to extract the leading low temperature behavior of the Lyapunov exponent and show that it saturates the bound on chaos.

Let us consider the regularised retarded OTOC for the impurity spin, defined as
\begin{equation}
\begin{split}
    C_{S}(t_{1},t_{2}) &= \frac{\theta(t_{1}) \theta(t_{2})}{N^{4}} \sum_{\alpha,\beta,\gamma,\delta = 1}^{N} \textrm{Tr} \Big[ \sqrt{\rho} \big[ \hat{S}_{\alpha \beta}(t_{1}),\hat{S}_{\delta \gamma}(0)  \big] \\
    &\times \sqrt{\rho} \big[ \hat{S}_{\alpha \beta}(t_{2}),\hat{S}_{ \delta \gamma }(0)  \big]^{\dagger}  \Big]
\end{split} 
\end{equation}
First we need to rewrite this correlator as an average over the two-fold Keldysh action. We perform the Keldysh rotation for the spin degrees of freedom and define
\begin{equation}
    S^{\sigma,cl}_{\alpha \beta} = \frac{S^{\sigma,+}_{\alpha \beta} + S^{\sigma,-}_{\alpha \beta}}{\sqrt{2}}, \qquad S^{\sigma,q}_{\alpha \beta} = \frac{S^{\sigma,+}_{\alpha \beta} - S^{\sigma,-}_{\alpha \beta}}{\sqrt{2}} 
\end{equation}
with $S_{\alpha \beta}^{\sigma, \pm} = \overline{f}_{\alpha}^{\sigma,\pm} f_{\beta}^{\sigma,\pm}$. Then we can rewrite  $C_{S}$ as 
\begin{equation}
    C_{S}(t_{1},t_{2}) = - \frac{1}{N^{4}}  \sum_{\alpha,\beta,\gamma,\delta } \left\langle S_{\alpha \beta}^{d,cl}(t_{1}) S_{\delta \gamma}^{d,q}(0) S_{\beta \alpha}^{u,cl}(t_{2}) S_{\gamma \delta}^{u,q}(0) \right\rangle \, .
\end{equation}
It will be convinient to express $S^{\sigma,cl}$ and $S^{\sigma,q}$ in terms of $f^{\sigma,cl}$ and $f^{\sigma,q}$:

\begin{equation}
    S_{\alpha \beta}^{\sigma,cl} = \overline{\Psi}_{\alpha}^{\sigma} \gamma^{q} \Psi_{\beta}^{\sigma} \qquad S_{\alpha \beta}^{\sigma,q} = \overline{\Psi}_{\alpha}^{\sigma} \gamma^{cl} \Psi_{\beta}^{\sigma}   
\end{equation}
where
\begin{equation}
    \Psi^{\sigma} = 
\begin{pmatrix}
    f^{\sigma,cl} \\
    f^{\sigma,q}
\end{pmatrix},
\qquad
\overline{\Psi}^{\sigma} =
\begin{pmatrix}
    \overline{f}^{\sigma,cl} & \overline{f}^{\sigma,q}
\end{pmatrix},
\end{equation}
and where
\begin{equation}
\gamma^{cl} = 
\begin{pmatrix}
    1 & 0 \\
    0 & 1
\end{pmatrix},
\qquad
\gamma^{q} = 
\begin{pmatrix}
    0 & 1 \\
    1 & 0
\end{pmatrix}
\end{equation}
The impurity spin OTOC now takes the form
\begin{widetext}
\begin{equation}
\begin{split}
    C_{S}(t_{1},t_{2}) &= - \frac{1}{N^{4}} \sum_{\alpha,\beta,\gamma,\delta = 1}^{N} \big\langle \overline{\Psi}_{\alpha}^{d}(t_{1}) \gamma^{q} \Psi_{\beta}^{d}(t_{1}) \overline{\Psi}_{\delta}^{d}(0) \gamma^{cl} \Psi_{\gamma}^{d}(0) \overline{\Psi}_{\beta}^{u}(t_{2}) \gamma^{q} \Psi_{\alpha}^{u}(t_{2}) \overline{\Psi}_{\gamma}^{u}(0) \gamma^{cl} \Psi_{\delta}^{u}(0)    \big\rangle \\
    &= - \sum_{  \substack{\mu_{1} \cdots \mu_{4} \\ \nu_{1} \cdots \nu_{4}}   }  \gamma^{q}_{\mu_{1} \mu_{2}} \gamma^{cl}_{\nu_{2} \nu_{1}} \gamma^{q}_{\mu_{4} \mu_{3} } \gamma^{cl}_{\nu_{3} \nu_{4} } C_{\mu_{1} \nu_{1} ; \mu_{2} \nu_{2}}^{\mu_{3}  \nu_{3} ; \mu_{4}  \nu_{4}}(t_{1},t_{2}) \, .
\end{split}
\end{equation}

with
\begin{equation}
    C_{\mu_{1} \nu_{1} ; \mu_{2} \nu_{2}}^{\mu_{3}  \nu_{3} ; \mu_{4}  \nu_{4}}(t_{1},t_{2}) = \frac{1}{N^{4}} \sum_{\alpha,\beta,\gamma,\delta = 1}^{N} \left\langle \overline{f}_{\alpha}^{d,\mu_{1}}(t_{1}) f_{\beta}^{d,\mu_{2}}(t_{1}) \overline{f}_{\delta}^{d,\nu_{2}}(0) f_{\gamma}^{d,\nu_{1}}(0) \overline{f}_{\beta}^{u,\mu_{4}}(t_{2}) f_{\alpha}^{u,\mu_{3}}(t_{2}) \overline{f}_{\gamma}^{u,\nu_{3}}(0) f_{\delta}^{u,\nu_{4}}(0)  \right\rangle 
\end{equation}

\end{widetext}
As previously, we expand the correlator $C_{\mu_{1} \nu_{1} ; \mu_{2} \nu_{2}}^{\mu_{3}  \nu_{3} ; \mu_{4}  \nu_{4}}(t_{1},t_{2})$ in powers of the interaction vertex and keep only the leading diagrams in the large $N$ and $K$ limit, which are shown on Fig. \ref{fig7}. Remarkably, these diagrams can be factorized into the product of auxiliary fermions OTOC
\begin{equation}\label{eqn:wick_OTOC_Spin}
    C_{\mu_{1} \nu_{1} ; \mu_{2} \nu_{2}}^{\mu_{3}  \nu_{3} ; \mu_{4}  \nu_{4}}(t_{1},t_{2}) = \overline{F}_{\mu_{1} \nu_{1}}^{\mu_{3} \nu_{3}}(t_{1},t_{2}) F_{\mu_{2} \nu_{2}}^{\mu_{4} \nu_{4}}(t_{1},t_{2}) \, ,
\end{equation}
where
\begin{align}
    F_{\mu_{2} \nu_{2} }^{\mu_{4} \nu_{4}}(t_{1},t_{2}) &= \frac{1}{N^{2} } \sum_{\beta , \delta} \big\langle f_{ \beta}^{\mu_{2},d}(t_{1}) \overline{f}_{ \delta}^{\nu_{2}, d}(0) \overline{f}_{ \beta}^{\mu_{4},u}(t_{2}) f_{ \delta}^{\nu_{4},u}(0) \big\rangle \\
    \overline{F}_{\mu_{1} \nu_{1} }^{\mu_{3} \nu_{3}}(t_{1},t_{2}) &= \frac{1}{N^{2} } \sum_{\alpha , \gamma} \big\langle \overline{f}_{\alpha}^{\mu_{1},d}(t_{1}) f_{\gamma}^{\nu_{1}, d}(0) f_{\alpha}^{\mu_{3},u}(t_{2}) \overline{f}_{\gamma }^{\nu_{3},u}(0) \big\rangle 
\end{align}

Using the expressions of the matrices $\gamma^{cl}$ and $\gamma^{q}$, the fact that the correlator $ \langle f^{q} \overline{f}^{cl} \rangle $ is zero and symetries between the different OTOCs we can simplify the sum into
\begin{equation}
\begin{split}
    C_{S}(t_{1},t_{2}) &= - \Big[ F_{+ +}(t_{1},t_{2}) F_{- -}(t_{1},t_{2})^{*} \\
    &- F_{+ -}(t_{1},t_{2}) F_{- +}(t_{1},t_{2})^{*}  + \textrm{c.c} \Big]
\end{split}
\end{equation}
with
\begin{equation}
\begin{split}
    F_{s s^{\prime}}(t_{1},t_{2}) &\propto \frac{1}{N^{2}} \sum_{\alpha,\beta = 1}^{N} \textrm{Tr} \bigg[ \sqrt{\rho} \Big[ \hat{f}_{\alpha}(t_{1}), \hat{f}_{\beta}^{\dagger}(0) \Big]_{s}\\
    &\times \sqrt{\rho} \Big[ \hat{f}_{\alpha}(t_{2}), \hat{f}_{\beta}^{\dagger}(0) \Big]_{s^{\prime}}^{\dagger} \bigg]
\end{split}
\end{equation}
with $ s,s^{\prime} = \pm $. The four possible combinations of product of commutator and anticommutator $F_{s s^{\prime}}$ appear in the definition of the OTOC. We have already computed the first one $F_{+ +}$ and one could be tempted to write down a ladder equation for the other three and try to find the scrambling rate by the same technique that we have used above. However the Keldysh structure of the resulting Bethe-Salpeter equations makes it hard to find a closed analytic solution. Since we are only interested in computing the scrambling rate of the spin, we will take a different approach. To simplify the discussion from now on we drop the spin and channel indices and we take $t_{1} = t_{2} = t$. Then we can write
\begin{equation}\label{eqn:C_S}
    C_{S}(t) = - 4 \, \theta(t) \, \textrm{Re} \Big\{ \textrm{Tr} \Big[ \sqrt{\rho} \hat{f}(t)\hat{f}^{\dagger}(0) \sqrt{\rho} \hat{f}^{\dagger}(t)\hat{f}(0)  \Big]^{2} \Big\} + \cdots
\end{equation}
Here we have only shown the OTOC part of $C_{S}$ which is responsible for the exponential chaotic growth. To see how it behaves, notice that
\begin{equation}
\begin{split}
    \textrm{Tr} \Big[ \sqrt{\rho} &\hat{f}(t) \hat{f}^{\dagger}(0) \sqrt{\rho} \hat{f}^{\dagger}(t)\hat{f}(0)  \Big] \\
    &= \textrm{Tr} \Big[ \rho^{\frac{1}{4}} \hat{f}\Big( t + i \frac{\beta}{4} \Big) \rho^{\frac{1}{4}} \hat{f}^{\dagger}(0) \rho^{\frac{1}{4}} \hat{f}^{\dagger}\Big( t + i \frac{\beta}{4} \Big) \rho^{\frac{1}{4}} \hat{f}(0)  \Big] \\
    &= F_{1 / 4} \Big( t + i \frac{\beta}{4}, t + i \frac{\beta}{4} \Big)
\end{split}
\end{equation}
with 
\begin{equation}
    F_{1/ 4}(t_{1},t_{2}) \equiv \textrm{Tr} \Big[ \rho^{\frac{1}{4}} \hat{f}( t_{1}) \rho^{\frac{1}{4}} \hat{f}^{\dagger}(0) \rho^{\frac{1}{4}} \hat{f}^{\dagger}( t_{2} ) \rho^{\frac{1}{4}} \hat{f}(0)  \Big]
\end{equation}
This function $F_{1/4}$, which corresponds to a different OTOC regularisation~\cite{tsujii2018bound,romero2019which}, satisfies the same ladder equation as the retarded OTOC
\begin{equation}
\begin{split}
    F_{1 / 4}(t_{1},t_{2}) &= F_{1 / 4}^{0}(t_{1},t_{2}) + \int \,d t_{3} d t_{4} \, K(t_{1},t_{2},t_{3},t_{4}) \\
    &\times F_{1 / 4}(t_{3},t_{4})
\end{split}
\end{equation}
with $K(t_{1},t_{2},t_{3},t_{4})$ the kernel defined above. Thus $F_{1 / 4} (t_{1},t_{2})$ is of the form 
\begin{equation}
    F_{1 / 4}(t_{1},t_{2}) = e^{i \frac{\alpha}{\beta} (t_{1} - t_{2})} f(t_{1} - t_{2}) e^{\lambda_{L} \frac{t_{1} + t_{2}}{2}}
\end{equation}
with $f$ a \textit{real} function (the expression is written above). Then we conclude that $F_{1 / 4}$ grows exponentially with a Lyapunov exponent $\lambda_L$ equal to the one of the retarded OTOC correlator. 
\begin{equation}
    F_{1 / 4} \Big( t + i \frac{\beta}{4}, t + i \frac{\beta}{4} \Big) = f(0) \, e^{i \frac{\beta \lambda_{L}}{4}}  e^{\lambda_{L} \frac{t_{1} + t_{2}}{2}}
\end{equation}
Using this result and the expression for the spin OTOC in Eq.(\ref{eqn:C_S}) we obtain 
\begin{equation}
C_{S}(t) \sim f(0) \cos\Big( \frac{\beta \lambda_{L}}{2} \Big) e^{2 \lambda_{L} \frac{t_{1} + t_{2}}{2}}
\end{equation}
We conclude therefore that the impurity spin OTOC has a Lyapunov exponent given by \emph{twice} the value of the fermionic and bosonic one. This factor of two is crucial since it allows the impurity spin, a physical and gauge invariant degree of freedom of our theory, to become maximally chaotic and to saturates the bound on chaos for $\gamma=1$, i.e. when the size of the impurity spin is equal to the number of channels. Right at this point bosons and fermions, the composite degrees of freedom of the impurity spin are also chaotic, but only acquire half of the maximal Lyapunov exponent,  a sort of equipartition of maximal chaos.

\section{Discussion}\label{sec:discussion}

The results obtained in the previous section show that the MCK model in the infinite $N,K$ limit at fixed $K/N$ is maximally chaotic and saturates the bound on chaos. This is a striking result from several standpoints that we briefly highlight here. First, maximal chaos occurs in a model that, quite differently from the SYK and its variations, does not present any random all-to-all coupling but only boundary interactions between a large quantum spin, strongly coupled to a large number of free fermionic channels. As we will argue this has direct consequences on the physical origin of the fast scrambling dynamics of the MCK model that sets it apart from other well known examples. 

A second reason for which this result is remarkable is from the point of view of the properties of the MCK model. Infact these models are integrable~\cite{tsvelik1984exact,jerezPRB1998} at any finite $N,K$. For example, for $N=K=2$ the model reduces to the two-channel Kondo model (2CK) which is known to be integrable at low energy. In particular it is known from CFT that 
the spectrum is made of a finite number of tower of states, where within each tower is the levels are equally spaced by $\pi v_F/L$. Hence, although the system is a non-Fermi Liquid, there is a quasiparticle description which is revealed by CFT~\cite{affleck1991universal,georges1994emery,georges1995solution}.
Furthermore in this case at zero temperature there is a residual entropy coming from a decoupled Majorana mode~\cite{emery1992mapping}. Recent work investigating the OTOC in the 2CK model~\cite{dora2017information} did not find any trace of exponential growth and Lyapunov chaos, a result which is compatible with the fact that models with finite local Hilbert space (i.e. spin chains) do not have an exponential regime for the growth of OTOC.

To reconcile the known results at finite $N,K$ with our results in the limit $N,K\rightarrow\infty$ is it crucial to understand how the finite-size spectrum, residual entropy and OTOC evolve in the MCK as the large $N,K$ limit is approached. The case of the SYK model in this respect is well understood: the finite-size spectrum contains exponentially many states above the groundstate, with exponentially small level spacing, leading to an extensive residual zero temperature entropy~\cite{chowdhury2022sachdev} if the large volume limit is taken first with respect to the zero-temperature one.
However the structure of the spectrum for the finite $N,K$ MCK model is very different from the case of SYK. In the 2CK model the residual entropy at $T=0$ arises from high-energy, large quantum number states whose level spacing is still regular and set by the inverse system size. In the thermodynamic limit, as the level spacing becomes small, these high-energy states with high degeneracy yield a residual entropy, when temperature is sent to zero after the taking the 
thermodynamic limit~\cite{affleck1991universal}.
A similar mechanism is at play for any $K,N$ finite, with levels described by $SU_K(N)$. We can now understand qualitatively how the structure of the spectrum of the MCK model evolves as $N,K$ are increased: while the CFT tower with the $1/L$ spacing is always there, the key is the behavior of the spectrum at large quantum numbers and whether exponentially many-levels contribute. This interpretation is indeed confirmed by looking at the residual entropy of the MCK for generic $N,K$, computed in Ref.~\cite{parcollet1998overscreeneed}, which gives a result of order $N$  in the large $N$ limit, where $N$ is the size of the spin. We conjecture that the fast scrambling we find in our calculation is associated with this manifold of states with high-degeneracies.


What are the consequences for the OTOC at finite but large $N,K$? The exponential growth of the OTOC is expected to hold on time scales shorter than the Ehrenfest time~\cite{maldacena2016abound}, that in the present case should be identified with $t_E\sim \mbox{log}N$. In the case of the SYK the long-time dynamics of the OTOC is controlled by Goldstone-mode fluctuations associated to time-reparametrization. These fluctuations act as a restoring force giving rise to a time scale above which a power-law decay of the OTOC is observed~\cite{Bagrets_2016,Bagrets_2017}. 
We can speculate that a mechanism similar to this one will be at play in controlling the long time behavior of the OTOC for $t\gg t_E$ at finite $N,K$.

\section{Conclusions}
\label{sec:conclusions}

Models that saturate the quantum bound on chaos are appealing for a wide range of communities, from high-energy theory to condensed matter and statistical mechanics. While few examples displaying maximal chaos exist, beyond the celebrated SYK model, they typically involve extended systems with random all to all interactions or conformal symmetries. In this work we exhibit arguably the simplest model of fast scrambler: a quantum $SU(N)$ spin coupled to a metallic bath with $K$ channels, in the large $N$ and $K$ limit, with fixed $\gamma = K / N$. In this limit the quantum spin fractionalise due to strong correlations into auxiliary fermionic and bosonic degrees of freedom, coupled to the conduction electrons of the metallic bath. Using the generalised two-fold Keldysh time contour, we have computed diagrammatically the OTOC for auxiliary fermions and bosons as well as the conduction electrons. We have demonstrated that for all these degrees of freedom the Lyapunov exponent is linear in temperature as $T\rightarrow0$, with a tunable prefactor maximum at $\gamma= 1 $. Finally, we have shown that the OTOC of the physical impurity spin factorizes in the large $N$ limit and can be deduced from the OTOC of the auxiliary fermions. In particular for the value $\gamma = K/N= 1$ the spin OTOC saturates the bound on chaos.  

The onset of maximal chaos raises the intriguing question of whether there is a gravity analogue of the MCK model, i.e. whether the effective theory of the soft mode associated to a broken reparametrization symmetry would be related to some model of gravity. Remarkably, holographic realization of Multichannel Kondo impurity models have been proposed in the context of the AdS/CFT correspondence~\cite{muck2011polyakov,erdmenger2013holographic,erdmenger2017holographic}, even though their chaos properties on the gravity side has not been investigated to the best of our knowledge. 

The field theory technique used in this work to evaluate the OTOC and Lyapunov exponent of composite operators such as the impurity spin could find immediate applications in the investigation of scrambling and chaos in models for doped quantum spin glasses which display similar fractionalisation \cite{joshi2020deconfined,christos2022critical,christos2022spin}.
Finally, a natural extension of the present work would be to investigate the non-equilibrium properties of this model and compare it to the known results on the SYK model. In particular, it could be interesting to see if the relaxation towards thermal equilibrium in this model also displays a Planckian behaviour, with a thermalization rate proportional to the temperature.

\section*{Acknowledgments}

This work was supported by the ANR grant "NonEQuMat" (ANR-19-CE47-0001). 

\appendix

\widetext

\section{Details on Double Contour}\label{app:Keldysh_double}

In this appendix we provide additional details on the two-fold Keldysh technique used in the main text to compute the retarded regularized OTOCs, like the $f$-OTOC
\begin{equation}
    C_{f}(t_{1},t_{2}) = \frac{\theta(t_{1}) \theta(t_{2})}{N^{2} } \sum_{\alpha , \beta = 1}^{N} \textrm{Tr} \Big[ \sqrt{\rho} \big\{ \hat{f}_{\alpha}(t_{1}),\hat{f}_{\beta}^{\dagger}(0)  \big\}  \sqrt{\rho} \big\{ \hat{f}_{\alpha}(t_{2}),\hat{f}_{ \beta}^{\dagger}(0)  \big\}^{\dagger}  \Big] \, ,
\end{equation}
where $\sqrt{\rho} = e^{- \beta \hat{H} / 2} $ and $ \hat{f}_{\alpha}(t) = e^{i \hat{H} t} \hat{f}_{\alpha} e^{- i \hat{H}t}$. By rearranging the exponential factors, $C_{f}(t_{1},t_{2})$ can be rewritten like
\begin{equation}
    C_{f}(t_{1}, t_{2}) = \frac{\theta(t_{1}) \theta(t_{2})}{N^{2}} \sum_{\alpha,\beta = 1}^{N} \left\langle \Big\{ \Hat{f}_{\alpha}\Big(t_{1} - i \frac{\beta}{2}\Big), \Hat{f}_{\beta}^{\dagger}\Big(- i \frac{\beta}{2}\Big) \Big\} \Big\{ \Hat{f}_{\alpha}(t_{2}), \Hat{f}_{\beta}^{\dagger}(0) \Big\}^{\dagger} \right\rangle_{\beta} \, ,
    \label{eq:OTOC_C_f}
\end{equation}
with $\langle \cdots \rangle_{\beta} = \textrm{Tr}[ \cdots e^{- \beta \hat{H}}] / Z$ the thermal average. The partition function of the model $Z$ along the two-fold time contour of Fig. \ref{fig2} is given by

\begin{equation}
     Z = \int \mathcal{D} [\overline{f},f,\overline{c},c,\overline{B},B,\mu] e^{i (S_{0} + S_{int})  }
\end{equation}
with $S_{0}$ the non-interacting action and $ S_{int}$ the interacting part 

\begin{equation}
    S_{int} = - \frac{i}{\sqrt{N}} \int_{- \infty}^{+\infty} d t \sum_{\sigma = u ,d} \sum_{s = +,-} \sum_{i \alpha} \Big( \overline{B}_{i}^{\sigma,s}(t) \overline{c}^{\sigma,s}_{i \alpha}(t) f_{\alpha}^{\sigma,s}(t) + B^{\sigma,s}_{i}(t) \overline{f}^{\sigma,s}_{\alpha}(t) c^{\sigma,s}_{i \alpha}(t) \Big) 
\end{equation}
$\sigma = u,d$ denotes the upper and lower folds of the contour and $+,-$ the forward and backward branch on the corresponding fold. Thermal expectation values $\langle \cdots \rangle_{\beta}$ can then be represented as averages over the action $S$ on the two-fold contour $\langle \cdots \rangle_{S}$, like for instance
\begin{equation}
    \Big\langle \hat{f}\Big( t_{1} - i \frac{\beta}{2} \Big) \hat{f}^{\dagger}\Big( - i \frac{\beta}{2} \Big)  \hat{f}^{\dagger}(t_{2}) \hat{f}(0) \Big\rangle_{\beta} = \Big\langle f^{d,-}(t_{1}) \overline{f}^{d,+}(0) \overline{f}^{u,-}(t_{2}) f^{u,+}(0) \Big\rangle_{S} \, .
    \label{eq:two_fold_correlator}
\end{equation}
The retarded OTOCs, like \ref{eq:OTOC_C_f}, are more easily represented by using the Keldysh rotated fields on each time fold $ \sigma $
\begin{align}
    f_{\alpha}^{\sigma, cl/q}(t) &= \frac{ f_{\alpha}^{\sigma, +}(t) \pm f_{\alpha}^{\sigma, -}(t) }{\sqrt{2}}, \qquad \qquad
    \overline{f}_{\alpha}^{\sigma, cl/q}(t) = \frac{ \overline{f}_{\alpha}^{\sigma, +}(t) \mp \overline{f}_{\alpha}^{\sigma, -}(t) }{\sqrt{2}} \label{eq:Keldysh_rotation_f} \\
    B_{i}^{\sigma,cl/q}(t) &= \frac{ B_{i}^{\sigma,+}(t) \pm B_{i}^{\sigma,-}(t)  }{\sqrt{2}}, \qquad \qquad
    \overline{B}_{i}^{\sigma,cl/q}(t) = \frac{ \overline{B}_{i}^{\sigma,+}(t) \pm \overline{B}_{i}^{\sigma,-}(t)   }{\sqrt{2}} \\
    c_{i \alpha}^{\sigma, cl/q}(t) &= \frac{ c_{\alpha}^{\sigma, +}(t) \pm c_{i \alpha}^{\sigma, -}(t) }{\sqrt{2}}, \qquad \qquad
    \overline{c}_{i \alpha}^{\sigma, cl/q}(t) = \frac{ \overline{c}_{i \alpha}^{\sigma, +}(t) \mp \overline{c}_{i \alpha}^{\sigma, -}(t) }{\sqrt{2}} 
\end{align}

In this basis, the interacting part of the action on the two-folds Keldysh contour is expressed as

\begin{equation}
\begin{split}
    S_{int} = - \frac{i}{\sqrt{2 N}} \int_{- \infty}^{+ \infty} d t \sum_{i,\alpha,\sigma} \Big[ &\overline{B}_{i}^{\sigma,cl} \big( \overline{c}_{i \alpha}^{\sigma,cl} f_{\alpha}^{\sigma,cl} + \overline{c}_{i \alpha}^{\sigma,q} f_{\alpha}^{\sigma,q}  \big) + \overline{B}_{i}^{\sigma,q} \big( \overline{c}_{i \alpha}^{\sigma,cl} f_{\alpha}^{\sigma,q} + \overline{c}_{i \alpha}^{\sigma,q} f_{\alpha}^{\sigma,cl} \big) \\
    &+ B_{i}^{\sigma,cl} \big( \overline{f}_{\alpha}^{\sigma,cl} c_{i \alpha}^{\sigma,cl} + \overline{f}_{\alpha}^{\sigma,q} c_{i \alpha}^{\sigma,q}  \big) + B_{i}^{\sigma,q} \big( \overline{f}_{\alpha}^{\sigma,cl} c_{i \alpha}^{\sigma,q} + \overline{f}_{\alpha}^{\sigma,q} c_{i \alpha}^{\sigma,cl}  \big)   \Big]
\end{split}
\end{equation}

The intra-fold two-point functions are the same as in the usual single-fold Keldysh formalism and are identical on the upper and lower folds:

\begin{equation}
\mathbf{G}_{f}^{\sigma \sigma}
=  
\begin{pmatrix}
     \, G^{R}_{f} & \, G^{K}_{f} \\
    0 & \, G^{A}_{f}
\end{pmatrix}
\qquad
\mathbf{G}_{c}^{\sigma \sigma}
=  
\begin{pmatrix}
     \, G^{R}_{c} & \, G^{K}_{c} \\
    0 & \, G^{A}_{c}
\end{pmatrix}
\qquad
\mathbf{G}_{B}^{\sigma \sigma} =
\begin{pmatrix}
     \, G^{K}_{B} &  \, G^{R}_{B} \\
     \, G^{A}_{B} & 0
\end{pmatrix}
\qquad
\sigma = u,d
\end{equation}

In these notations the matrix elements for a given field $\psi$ are $  [ \textbf{G}^{\sigma \sigma^{\prime}}_{\psi} ]^{s s^{\prime}}(t,t^{\prime}) = - i \xi \langle \psi^{\sigma,s}(t)  \, \overline{\psi}^{\sigma^{\prime},s^{\prime}}(t^{\prime}) \rangle $ with $\sigma,\sigma^{\prime} = u,d $ and $s,s^{\prime} = cl, q$ and where $\chi = 1$ (resp. $- 1$) if $\psi$ is fermionic (resp. bosonic). We have omitted the spin and channel indices to simplify notations. The retarded, advanced and Keldysh Green's functions are defined by

\begin{align}
   i G_{\psi}^{R}(t,t^{\prime}) &= \xi  \theta(t - t^{\prime}) \, \langle [  \hat{\psi} , \hat{\psi}^{\dagger}(t^{\prime}) ]_{\xi}  \rangle \\
    i G_{\psi}^{A}(t,t^{\prime}) &= - \xi  \theta(t^{\prime} - t) \, \langle [  \hat{\psi} , \hat{\psi}^{\dagger}(t^{\prime}) ]_{\xi}  \rangle \\
    i G_{\psi}^{K}(t,t^{\prime}) &=  \xi  \, \langle [  \hat{\psi} , \hat{\psi}^{\dagger}(t^{\prime}) ]_{- \xi}  \rangle
\end{align}
where the average is over the thermal density matrix and $[A,B]_{\xi} = A B + \xi B A$. For the inter-fold Green's functions, only the Keldysh component is non-zero
\begin{equation}
\mathbf{G}_{f}^{\sigma \overline{\sigma}} = 
\begin{pmatrix}
    0 & \, G^{K}_{f,\sigma \overline{\sigma}}(t,t^{\prime}) \\
    0 & 0
\end{pmatrix},
\quad 
\mathbf{G}_{c}^{\sigma \overline{\sigma}} = 
\begin{pmatrix}
    0 & \, G^{K}_{c,\sigma \overline{\sigma}}(t,t^{\prime}) \\
    0 & 0
\end{pmatrix},
\quad
\mathbf{G}_{B}^{\sigma \overline{\sigma}} =
\begin{pmatrix}
     \, G^{K}_{B,\sigma \overline{\sigma} }(t,t^{\prime}) & 0\\
    0 & 0
\end{pmatrix}
\end{equation}
where we used the notation $\overline{u} = d$ and reciprocally. These are the so-called Wightman correlators, which are defined as the thermal averages

\begin{equation}
     i G^{K}_{\psi,u d}(t,t^{\prime}) = 2 \xi \,  \Big\langle \hat{\psi} \Big(t - i \frac{\beta}{2} \Big)  \hat{\psi}^{\dagger}(t^{\prime})  \Big\rangle, \qquad
      i G^{K}_{\psi, d u}(t,t^{\prime}) = - 2 \xi \, \Big\langle \hat{\psi}^{\dagger} \Big(t^{\prime} - i \frac{\beta}{2} \Big)  \hat{\psi}(t)  \Big\rangle
\end{equation}

To write the OTOC \ref{eq:OTOC_C_f} as a four-point function on the two-fold contour \ref{fig2}, one can develop the product of the two anti-commutators and use the identity

\begin{equation}
    \theta(t) \Big( \hat{f}(t) \hat{f}^{\dagger}(0) + \hat{f}^{\dagger}(0) \hat{f}(t) \Big) = \frac{1}{2} \Big[ T \big( \hat{f}(t) \hat{f}^{\dagger}(0) \big) - \Tilde{T} \big( \hat{f}(t) \hat{f}^{\dagger}(0) \big) +  \hat{f}(t) \hat{f}^{\dagger}(0) +  \hat{f}^{\dagger}(0) \hat{f}(t) \Big] \, ,
\end{equation}
where $T$ and $\Tilde{T}$ are the time-ordering and anti-time ordering operations. Using identities like \ref{eq:two_fold_correlator} and the definition of the rotated fields \ref{eq:Keldysh_rotation_f}, after some rearrangements we finally get

\begin{equation}
    C_{f}(t_{1},t_{2}) = \frac{1}{N^{2} } \sum_{\alpha , \beta = 1}^{N} \big\langle f_{ \alpha}^{d,cl}(t_{1}) \overline{f}_{ \beta}^{d,cl}(0) \overline{f}_{ \alpha}^{u,q}(t_{2}) f_{ \beta}^{u,q}(0) \big\rangle
\end{equation}

\section{Real time Green's function}\label{app:realtimeGF}

For bosons and fermions, the real time retarded, advanced, and Wightman two-point functions are related to the imaginary time Matsubara Green's function by analytic continuation

\begin{align}
    G^{R}(t) &= i \theta(t) \Big( G(\tau = it + \epsilon) - G(\tau = it - \epsilon) \Big) \, , \\
    G^{A}(t) &= - i \theta(- t) \Big( G(\tau = it + \epsilon) - G(\tau = it - \epsilon) \Big) \, , \\
    G^{K}_{d u}(t) &= 2 i G \Big( \tau = it + \frac{\beta}{2} \Big) \, , \\
    G^{K}_{ u d }(t) &= 2 i G \Big(\tau = it - \frac{\beta}{2} \Big) \, .
\end{align}
By using the conformal expression of the Matsubara Green's functions \ref{gf_mats}, one finds after some simple manipulations for the auxiliary fermions $f$:
\begin{align}
    G_{f}^{R}(t) &= - 2 i A_{f} \theta(t) \cos \Big( \pi \Delta_{f} - i \frac{\alpha}{2} \Big) \frac{e^{i \frac{\alpha}{\beta} t}}{\cosh \big( \frac{\alpha}{2} \big) } \bigg( \frac{\pi}{\beta \sinh \big( \frac{\pi t}{\beta} \big) } \bigg)^{2 \Delta{f}} \\
    G_{f}^{A}(t) &= 2 i A_{f} \theta(- t) \cos \Big( \pi \Delta_{f} + i \frac{\alpha}{2} \Big) \frac{e^{i \frac{\alpha}{\beta} t}}{\cosh \big( \frac{\alpha}{2} \big) } \bigg( \frac{\pi}{\beta \sinh \big( -\frac{\pi t}{\beta} \big) } \bigg)^{2 \Delta{f}} \\
    G_{f,du}^{K}(t) &= - 2 i A_{f} \frac{e^{i \frac{\alpha}{\beta} t}}{\cosh \big( \frac{\alpha}{2} \big) } \bigg( \frac{\pi}{\beta \cosh \big( \frac{\pi t}{\beta} \big) } \bigg)^{2 \Delta{f}} \\
    G_{f,u d}^{K}(t) &= 2 i A_{f} \frac{e^{i \frac{\alpha}{\beta} t}}{\cosh \big( \frac{\alpha}{2} \big) } \bigg( \frac{\pi}{\beta \cosh \big( \frac{\pi t}{\beta} \big) } \bigg)^{2 \Delta{f}} 
\end{align}
For the auxiliary bosons $B$, we have
\begin{align}
    G_{B}^{R}(t) &= - 2  A_{B} \theta(t) \sin \Big( \pi \Delta_{B} - i \frac{\alpha}{2} \Big) \frac{e^{i \frac{\alpha}{\beta} t}}{\cosh \big( \frac{\alpha}{2} \big) } \bigg( \frac{\pi}{\beta \sinh \big( \frac{\pi t}{\beta} \big) } \bigg)^{2 \Delta{B}} \\
    G_{B}^{A}(t) &= - 2 A_{B} \theta(- t) \sin \Big( \pi \Delta_{B} + i \frac{\alpha}{2} \Big) \frac{e^{i \frac{\alpha}{\beta} t}}{\cosh \big( \frac{\alpha}{2} \big) } \bigg( \frac{\pi}{\beta \sinh \big( -\frac{\pi t}{\beta} \big) } \bigg)^{2 \Delta{B}} \\
    G_{B,du}^{K}(t) &= - 2 i A_{B} \frac{e^{i \frac{\alpha}{\beta} t}}{\cosh \big( \frac{\alpha}{2} \big) } \bigg( \frac{\pi}{\beta \cosh \big( \frac{\pi t}{\beta} \big) } \bigg)^{2 \Delta{B}} \\
    G_{B,u d}^{K}(t) &= - 2 i A_{B} \frac{e^{i \frac{\alpha}{\beta} t}}{\cosh \big( \frac{\alpha}{2} \big) } \bigg( \frac{\pi}{\beta \cosh \big( \frac{\pi t}{\beta} \big) } \bigg)^{2 \Delta{B}} 
\end{align}
Finally, we get for the conduction electrons
\begin{align}
    G_{c,d u}^{K}(t) &= - 2 i \rho_{0} \frac{\pi}{\beta  \cosh \big(  \frac{\pi t}{\beta} \big) } \\
    G_{c,u d}^{K}(t) &=  2 i \rho_{0} \frac{\pi}{\beta  \cosh \big(  \frac{\pi t}{\beta} \big) }
\end{align}

\twocolumngrid


%

\end{document}